\documentclass[aps,pra,reprint,superscriptaddress,showpacs,floatfix,preprintnumbers,showkeys]{revtex4-1}

\usepackage{graphicx}
\usepackage{color}
\usepackage{amssymb}
\usepackage{amsmath}
\usepackage{dcolumn}
\usepackage{slashed}
\usepackage[american]{babel}
\usepackage[colorlinks,breaklinks=true,urlcolor=blue,citecolor=blue,linkcolor=blue,pdfstartview=FitH,pdfpagemode=UseNone]{hyperref}
\DeclareGraphicsExtensions{.pdf}

\usepackage[T1]{fontenc}
\usepackage{txfonts}

\newcommand{\headerstyle}[1]{{\bf{#1}}---}                  
\newcommand{\abs}[1]{\lvert #1 \rvert}                      
\newcommand{\wf}{\phi}                                      
\newcommand{\oppositeindex}[1]{\slashed{#1}}                
\newcommand{\chempot}{\mu}                                  
\newcommand{\osclength}{a_{\mathrm{osc}}}                   
\newcommand{\intmeasure}{\mathrm{d}^3 r}                    
\newcommand{\intra}[1]{\alpha_{#1}}                         
\newcommand{\inter}[1]{\beta_{#1}}                          
\newcommand{\determinant}{D}                                
\newcommand{\peakdensity}[1]{X_{#1}}                        
\newcommand{\formIIgrad}[2]{Y_{#1 #2}}                      
\newcommand{\secondmoment}[2]{\langle {#2}^2 \rangle_{#1}}  
\newcommand{\momentratio}[3]{\delta^{#1}_{{#2}{#3}}}        
\newcommand{\estimate}{(c_{AB}/c_{B})_{\mathrm{est}}}       
\newcommand{\outerellipsoid}{\mathcal{S}}                   
\newcommand{\closure}[1]{\bar{#1}}                          

\begin{document}
\title{Determining the interspecies interaction strength of a two-species Bose--Einstein condensate from the density profile of one species}
\date{May 1, 2018}
\author{P. Kuopanportti}\affiliation{Department of Physics, University of Helsinki, P.O. Box 43, FI-00014 Helsinki, Finland}
\author{Y. M. Liu}\affiliation{Department of Physics, Shaoguan University, Shaoguan, 512005, China}
\author{Y. Z. He}\affiliation{School of Physics, Sun Yat-Sen University, Guangzhou, 510275, China}
\author{C. G. Bao}\affiliation{School of Physics, Sun Yat-Sen University, Guangzhou, 510275, China}

\begin{abstract}
We study harmonically trapped two-species Bose--Einstein condensates within the Gross--Pitaevskii formalism. By invoking the Thomas--Fermi approximation, we derive an analytical solution for the miscible ground state in a particular region of the system's parameter space. This solution furnishes a simple formula for determining the relative strength of the interspecies interaction from a measurement of the density distribution of only one of the two species. Accompanying numerical simulations confirm its accuracy for sufficiently large numbers of condensed particles. The introduced formula provides a condensate-based scheme that complements the typical experimental methods of evaluating interspecies scattering lengths from collisional measurements on thermal samples.
\end{abstract}
\pacs{03.75.Mn, 67.85.Fg, 03.75.Hh}
\keywords{Bose--Einstein condensation, Multicomponent condensate, Thomas--Fermi approximation}

\maketitle

\headerstyle{Introduction}Binary mixtures of Bose--Einstein condensates (BECs) have been extensively studied in recent years, both experimentally and theoretically. In experiments to date, these so-called two-species BECs have been produced by using either two different elements~\cite{Fer2002.PRL89.053202,Mod2002.PRL89.190404,Tha2008.PRL100.210402,Aik2009.NJP11.055035,Cat2009.PRL103.140401,Car2011.PRA84.011603,Ler2011.EPJD65.3,Pas2013.PRA88.023601,Wac2015.PRA92.053602,Wan2015.JPhysB49.015302}, two distinct isotopes of the same element~\cite{Pap2008.PRL101.040402,Sug2011.PRA84.011610,Ste2013.PRA87.013611}, or a single isotope in two different internal states~\cite{Mya1997.PRL78.586,Hal1998.PRL81.1539,Mat1999.PRL83.2498,Del2001.PRA63.051602,Sch2004.PRL93.210403,And2009.PRA80.023603}. Theoretical studies, in turn, have addressed diverse phenomena such as segregation~\cite{Tim1998.PRL81.5718,Ao1998.PRA58.4836,Ind2015.PRA91.033615,Roy2015.PRA92.011601,Pol2015.PRA91.053626} and the associated symmetry breaking~\cite{Chu1999.PRA59.1473,Tri2000.JPhysB33.4017,Rib2002.PRA65.063614,Svi2003.PRA67.053608,Gau2010.JPhysB43.095302}, wetting phase transitions~\cite{Sch2015.PRA91.013626}, and exotic vortex structures~\cite{Mue2002.PRL88.180403,Kas2004.PRL93.250406,Kas2005.PRA71.043611,Kas2005.IJMPB19.1835,Yan2008.PRA77.033621,Kas2009.PRA79.023606,Mas2011.PRA84.033611,Kuo2012.PRA85.043613,Kuo2015.PRA91.043605,Gal2015.NJP17.103040}, to name but a few.

A key ingredient that gives rise to these phenomena and sets the two-species system apart from the single-species BEC is, quite obviously, the interspecies interaction, which is taken here to be of the zero-range density--density type. It can have a drastic effect on the ground-state density distributions, leading, for example, to segregation of the two condensates when it is strongly repulsive~\cite{Pap2008.PRL101.040402,Car2011.PRA84.011603,Wan2015.JPhysB49.015302}. In this paper, we demonstrate how the ground-state shapes of the coupled condensates encode crucial information about the interspecies interaction, even when no phase separation occurs, and how the information can be conveniently extracted.
Specifically, based on the analytical Thomas--Fermi (TF) formalism, we derive below a simple formula [Eq.~\eqref{ratio2}] that can be used to determine the relative strength of the interspecies interaction from a measurement of the density distribution of just one of the two miscible condensate species.

\headerstyle{Gross--Pitaevskii model}As the starting point of our theoretical treatment, let $N_{A}$ bosonic atoms of species $A$ and mass $m_{A}$ and $N_{B}$ bosonic atoms of species $B$ and mass $m_{B}$ be confined and Bose--Einstein condensed in three-dimensional, concentric harmonic traps. Atoms within each species are assumed to interact through repulsive contact interaction of strength $c_{S}=4\pi\hbar^2 a_{SS}/m_S > 0$, where $S\in\left\{A,B\right\}$ and $a_{SS}$ is the $s$-wave scattering length between atoms of species $S$. The interspecies contact interaction strength $c_{AB}=2\pi\hbar^2 a_{AB}\bigl(m_{A}^{-1}+m_{B}^{-1}\bigr)$, where $a_{AB}$ is the positive or negative interspecies $s$-wave scattering length, is taken to be weak enough for the two species to remain miscible~\footnote{If the finite size and inhomogeneity of the system are neglected, the miscible regime corresponds to $\abs{c_{AB}} < \sqrt{c_{A}c_{B}}$.}. The concentric harmonic traps are written as $V_\mathrm{trap}^{S}\bigr(\mathbf{r})=m_{S}(\omega _{Sx}^{2}x^{2}+\omega _{Sy}^{2}y^{2}+\omega_{Sz}^{2}z^{2}\bigr)/2$, where the trap frequencies $\omega_{Sl}$, $l\in\left\{ x,y,z\right\}$, may all be different. It should be noted, however, that we have assumed the two traps to be co-aligned such that they can both be assigned the same symmetry axes (which we have selected as our Cartesian coordinate axes). For the sake of convenience and notational symmetry, we introduce a mass $m$ and a frequency $\omega$ and hereafter use $\hbar \omega $ and $\osclength\equiv \sqrt{\hbar/(m\omega )}$ as units of energy and length, respectively. Assuming that the temperature is close enough to zero, the ground state of the two-species BEC can be described accurately by the time-independent coupled Gross--Pitaevskii~(GP) equations~\cite{Esr1997.PRL78.3594,Pu1998.PRL80.1130,Ho1996.PRL77.3276} for the condensate wave functions $\wf_{S}$, $S\in\{A,B\}$:
\begin{equation}\label{eq:GP}
\begin{split}
\biggl[&-\frac{m}{2m_{S}}\nabla^{2}+\frac{1}{2}\left(\gamma_{Sx}x^{2}+\gamma_{Sy}y^{2}+\gamma _{Sz}z^{2}\right) \\&+\intra{S} \abs{\wf _{S}(\mathbf{r})}^2+\inter{S}\abs{\wf _{\oppositeindex{S}}(\mathbf{r})}^2-\chempot_{S}\biggr]\wf _{S}(\mathbf{r})
=0, 
\end{split}
\end{equation}
where $\gamma _{Sl}=m_S \omega _{Sl}^2/(m \omega^{2})$ is a dimensionless trap frequency, $\oppositeindex{S}$ is defined such that $\oppositeindex{A}=B$ and $\oppositeindex{B}=A$, the dimensionless coupling constants are
\begin{align}
\intra{S}&= 4\pi N_{S}  \frac{m  a_{SS}}{m_{S} \osclength},\\
\inter{S}&=2\pi N_{S} \frac{m\left(m_{A}+m_{B}\right)a_{AB}}{m_{A}m_{B}\osclength},
\end{align} 
and $\chempot _{S}$ are the chemical potentials that enter as Lagrange multipliers enforcing the unit normalizations $\int_{\mathbb{R}^3} \abs{\wf_{S}^{2}(\mathbf{r})}\intmeasure =1$. Since we will only consider flowless ground states, we can assume $\wf_{S} \in\mathbb{R}$.

\headerstyle{Thomas--Fermi solution}We introduce the TF approximation~(TFA)~\cite{Ho1996.PRL77.3276,Edw1995.PRA51.1382,Bay1996.PRL76.6}, which applies to sufficiently large numbers of condensed atoms and amounts to neglecting the kinetic energy terms. When both $\wf _{A}$ and $\wf _{B}$ are nonzero, the resulting TF versions of Eqs.~\eqref{eq:GP} can be written as
\begin{equation}\label{uv2}
\intra{S}\wf_{S}^2 + \inter{S} \wf_{\oppositeindex{S}}^2 = \chempot_{S}-\frac{1}{2}\left(\gamma _{Sx}x^{2}+\gamma _{Sy}y^{2}+\gamma _{Sz}z^{2}\right).
\end{equation}
If the determinant $\determinant\equiv \intra{A}\intra{B}-\inter{A}\inter{B}\neq 0$, we obtain
\begin{equation}\label{eq:formIIsol}
\wf _{S}^{2} =\peakdensity{S}-\formIIgrad{S}{x}x^{2}-\formIIgrad{S}{y}y^{2}-\formIIgrad{S}{z}z^{2},  
\end{equation}
where
\begin{align}
\peakdensity{S} &\equiv (\intra{\oppositeindex{S}}\chempot_{S}-\inter{S}\chempot_{\oppositeindex{S}})/\determinant,\label{eq:peakdensity} \\
\formIIgrad{S}{l} &\equiv (\intra{\oppositeindex{S}} \gamma _{Sl}-\inter{S}\gamma _{\oppositeindex{S}l})/(2\determinant).\label{eq:formIIgrad}
\end{align}
We will refer to the formal solution given by Eqs.~\eqref{eq:formIIsol} as Form~II. Equations~\eqref{eq:peakdensity} can be solved for the chemical potentials:
\begin{equation}\label{eq:chempot}
\chempot_{S} =\intra{S} \peakdensity{S}+\inter{S} \peakdensity{\oppositeindex{S}}.  
\end{equation}
The parameters $\formIIgrad{S}{l}$ defined in Eqs.~\eqref{eq:formIIgrad} are known once the input parameters are given, while $\peakdensity{S}$ remain unknown because they depend on $\chempot_{S}$.

If exactly one of the two wave functions, say $\wf_{A}$, is zero in a certain region of $\mathbb{R}^3$, the formal solution is
\begin{subequations}\label{eq:formIBsol}
\begin{align}
\wf _{A} &=0,  \label{eq:formIBsolA} \\
\wf _{B}^{2} &=\frac{1}{\intra{B}}\left[\chempot _{B}-\frac{1}{2}\left(\gamma
_{Bx}x^{2}+\gamma _{By}y^{2}+\gamma _{Bz}z^{2}\right)\right]. \label{eq:formIBsolB}
\end{align}
\end{subequations}
A solution of this type is referred to as Form~I$_{B}$, where the subscript $B$ indicates the nonvanishing species. Analogously, Form~I$_{A}$ can be defined. Together with the vacuum $\wf_{A}=\wf_{B}=0$, Forms~I$_{A}$, I$_{B}$, and II exhaust all possible types of local TF solutions of Eqs.~\eqref{eq:GP}.

If one of the wave functions in Form~II, say $\wf_A(\mathbf{r})$, reaches zero as we vary $\mathbf{r}$, we arrive at a boundary surface of Form~II (for instance, $\wf_{A}$ will reach zero upon increasing $x$ sufficiently if $\formIIgrad{A}{x}>0$). Crossing the boundary will lead to a transformation from Form~II to Form~I$_{B}$. It is emphasized that both wave functions are always continuous at the form boundaries; this is because the equations governing the two neighbouring forms become exactly the same for the boundary points. In this way the formal solutions, each with its own specific domain of definition, will be naturally and continuously linked up to form the complete piecewise-defined TF solution over entire $\mathbb{R}^3$. The complete TF wave functions, however, will not in general be differentiable at the form boundaries. The two unknowns $\chempot_{S}$ appearing in the entire solution can be obtained from the two additional equations $\int_{\mathbb{R}^3}\wf_{S}^2\,\intmeasure=1$ for normalization.

The parameter space of the two-species model is fairly high-dimensional: even after all the redundancies are removed, one must specify the values of at least nine independent parameters in order to fix all the coefficients in Eqs.~\eqref{eq:GP}. Partly for this reason, we will not develop the general TF solution any further in what follows. Instead, for our purposes, it is sufficient to consider a specific type of TF solution satisfying the following assumptions: (i)~The isosurfaces of $\wf_{A}^2$ are ellipsoids, and $\wf_{A}^2$ has its maximum at the origin. (ii)~$\wf_{B}^{2}>0$ if $\wf_{A}^2 > 0$. (iii)~The boundary surfaces of the two condensates do not have any points in common. A TF solution satisfying assumptions (i)--(iii) will approximate the ground state of the system in a particular region of the whole parameter space. Note that these assumptions are different for the two species and hence should be used as the criteria for assigning the two labels $A$ and $B$.

Due to assumptions~(i) and~(ii), the solution has Form~II at the origin, and it follows from Eqs.~\eqref{eq:formIIsol} that $\peakdensity{A}=\wf_{A}^2\left(\mathbf{r}=\mathbf{0}\right)>0$ and $\peakdensity{B}=\wf_{B}^2\left(\mathbf{r}=\mathbf{0}\right)>0$. Assumption~(i) also implies that all the three $\formIIgrad{A}{l}>0$. Let us refer to the region in which $\wf_{A}$ remains nonzero as the inner region, $\Omega_{\mathrm{in}}\equiv \left\{ (x,y,z)\in\mathbb{R}^3\mid \sum_{l} \formIIgrad{A}{l} l^2 < \peakdensity{A} \right\}$. We know from assumption~(ii) that the solution has Form~II in $\Omega_\mathrm{in}$  and from assumption~(iii) that $\wf_{B}^2$ remains positive on the boundary ellipsoid $\partial\Omega_{\mathrm{in}}$. As we cross $\partial \Omega_{\mathrm{in}}$ to the outside, the solution acquires Form~I$_{B}$ [Eqs.~\eqref{eq:formIBsol}]. Since all the three $\gamma_{Bl}$ are positive by definition, the isosurfaces of $\wf_{B}^2$  in Eq.~\eqref{eq:formIBsolB} are also ellipsoids, and $\wf _{B}^{2}$ will reach zero on the ellipsoid $\left\{ (x,y,z)\in\mathbb{R}^3\mid \sum _{l}\gamma_{Bl}l^{2}=2\chempot _{B} \right\}\equiv \outerellipsoid$. Because both wave functions vanish outside $\outerellipsoid$, it is the boundary surface of the whole two-species BEC. The region between $\partial\Omega_{\mathrm{in}}$ and $\outerellipsoid$ is referred to as the outer region and denoted by $\Omega_{\mathrm{out}}$.

The normalization $1=\int_{\mathbb{R}^3}\wf_{A}^2 \intmeasure =\int_{\Omega_{\mathrm{in}}}\wf_{A}^2 \intmeasure$ yields
\begin{equation}
\peakdensity{A}=\left(\frac{15\sqrt{\formIIgrad{A}{x}\formIIgrad{A}{y}\formIIgrad{A}{z}}}{8\pi}\right)^{2/5}.  \label{eq:peakdensityA}
\end{equation}
From the normalization  $1=\int_{\mathbb{R}^3}\wf_{B}^2 \intmeasure =\int_{\closure{\Omega}_{\,\mathrm{in}}\cup \,\closure{\Omega}_{\mathrm{out}}}\wf_{B}^2 \intmeasure$  and Eqs.~\eqref{eq:chempot} and~\eqref{eq:peakdensityA}, we obtain
\begin{equation}\label{eq:chempotBfinal}
\frac{\chempot_{B}^{5/2}}{15}= \frac{\prod_l \gamma_{Bl}^{1/2}}{16\sqrt{2}\pi}\left(\intra{B}+\frac{5}{2}\inter{B}+ \sum_{l}\frac{2\intra{B}\formIIgrad{B}{l}-\gamma_{Bl}}{4\formIIgrad{A}{l}}\right).
\end{equation}
We can further use Eqs.~\eqref{eq:peakdensity} and~\eqref{eq:chempot} to write down closed analytical expressions for the remaining unknowns $\peakdensity{B}=\left(\chempot _{B}-\inter{B} \peakdensity{A}\right)/\intra{B}$ and $\chempot_{A}=\left(D\peakdensity{A}+\inter{A}\chempot_{B}\right)/\intra{B}$. Thus, all the quantities involved in $\wf_{A}$ and $\wf_{B}$ have now been determined in terms of the model input parameters, and thereby the desired TF solution has been obtained.

In order for the solution to be self-consistent, it must satisfy the assumptions made in its design. Since we must necessarily have
\begin{equation}\label{eq:inequality1}
\formIIgrad{A}{x}>0,\quad \formIIgrad{A}{y} >0,\quad \formIIgrad{A}{z} >0,
\end{equation}
the requirement $\peakdensity{A} > 0$ is immediately satisfied by Eq.~\eqref{eq:peakdensityA}. By utilizing standard techniques of analytical minimization, we can cast the constraint $\wf_{B}^2(\mathbf{r}) > 0\ \forall \mathbf{r}\in\closure{\Omega}_{\mathrm{in}}$, where $\wf_{B}^2$ is given by Eq.~\eqref{eq:formIIsol}, as the inequality
\begin{equation}\label{eq:inequality2}
\frac{\peakdensity{B}}{\peakdensity{A}} >  \max\left\{ 0,\max_l \frac{\formIIgrad{B}{l}}{\formIIgrad{A}{l}}\right\}.
\end{equation}
Furthermore, assumption (iii) implies that $\wf_{B}^2$ as given by Eq.~\eqref{eq:formIBsolB} must be positive on $\partial \Omega_{\mathrm{in}}$, which in turn requires that
\begin{equation}\label{eq:inequality3}
\chempot_{B} > \frac{\peakdensity{A}}{2}\max_{l} \frac{\gamma_{Bl}}{\formIIgrad{A}{l}}.
\end{equation}
As long as the ten input parameters $\gamma_{Sl}$, $\intra{S}$, and $\inter{S}$ are chosen such that the inequalities~\eqref{eq:inequality1}--\eqref{eq:inequality3} are satisfied, the TF solution derived above is self-consistent. 

\begin{table*}[t]
\caption{\label{table:oblate-oblate}RMS values of the coordinates $x$ and $z$ for ground-state density distribution of a harmonically trapped, three-dimensional two-species BEC with $N_{B}/N_{A}=2.5$, $c_{AB}/c_{A}=1.032$, $\omega_{Bx}/\omega_{Ax}=1.5$, $\omega_{Az}/\omega_{Ax}=2$, $\omega_{Bz}/\omega_{Bx}=1.5$, and $\omega_{Ay}/\omega_{Ax}=\omega_{By}/\omega_{Bx}=1$. The RMS values are given for both the numerically obtained GP solution and the Thomas--Fermi approximation (TFA) derived in the text. The second-to-last column shows the estimate $\estimate$ for the relative interspecies interaction strength, which is obtained by evaluating the right-hand side of Eq.~\eqref{ratio2} for the numerically obtained $\wf_{A}$. The last column gives the relative error of $\estimate$. For the TFA, Eq.~\eqref{ratio2} is exact and yields the true value $c_{AB}/c_{B} = 0.8$. We have used the values $m_{B}/m_{A}=0.471$ and $c_{B}/c_{A}=1.29$, which correspond to species $A$ being  \textsuperscript{87}Rb and species $B$ being \textsuperscript{41}K. The first column shows the value of the intraspecies interaction strength for species $A$, $\intra{A} = N_{A} c_{A}/(\hbar\omega \osclength^3)=4\pi N_{A} a_{AA}/\osclength$, where $\osclength=\sqrt{\hbar/\left(m\omega\right)}$, $m=m_{A}$, and $\omega=\omega_{Ax}$. All lengths are given in units of $\osclength$. (If we use $\omega=2\pi\times 100$ Hz and $a_{AA}=99\,a_\mathrm{B}$, we obtain $\osclength \approx 1.1$~$\mu\textrm{m}$ and $\intra{A} \approx 0.061 \times N_{A}$.) All the listed states satisfy inequalities~\eqref{eq:inequality1}--\eqref{eq:inequality3}, rendering our TF solution self-consistent.}
\begin{ruledtabular}
\begin{tabular}{ddddddddddd}
 & \multicolumn{2}{c}{~~~~~~~~$\secondmoment{A}{x}^{1/2}$} &  \multicolumn{2}{c}{~~~~~~~~$\secondmoment{B}{x}^{1/2}$} &  \multicolumn{2}{c}{~~~~~~~~$\secondmoment{A}{z}^{1/2}$}  &  \multicolumn{2}{c}{~~~~~~~~$\secondmoment{B}{z}^{1/2}$} & \multicolumn{2}{c}{~~~~$\estimate$}\\
\cline{2-3} \cline{4-5} \cline{6-7} \cline{8-9} \cline{10-11}
\multicolumn{1}{l}{$\intra{A}$} & \multicolumn{1}{r}{Numer.} &  \multicolumn{1}{r}{TFA} &   \multicolumn{1}{r}{Numer.} &  \multicolumn{1}{r}{TFA} &  \multicolumn{1}{r}{Numer.} &  \multicolumn{1}{r}{TFA} &  \multicolumn{1}{r}{Numer.} &  \multicolumn{1}{r}{TFA} & \multicolumn{1}{r}{Value} &  \multicolumn{1}{r}{Error ($\%$)} \\
\hline
10^0 & 0.7100 &  0.5230 &  0.7618 &  0.5724 &  0.5020 &  0.1410 &  0.5998 &  0.4221 &  7.5553 &844.409\\
10^1 & 0.9312 &  0.8289 &  1.0170 &  0.9072 &  0.5605 &  0.2235 &  0.7880 &  0.6689 & -2.2950 &-386.881\\
10^2 & 1.4346 &  1.3136 &  1.4392 &  1.4378 &  0.6500 &  0.3543 &  1.0822 &  1.0602 &  0.3135 &-60.807\\
10^3 & 2.1923 &  2.0820 &  2.2500 &  2.2788 &  0.7574 &  0.5615 &  1.6809 &  1.6803 &  0.6742 &-15.725\\
10^4 & 3.3380 &  3.2997 &  3.5994 &  3.6117 &  0.9883 &  0.8899 &  2.6607 &  2.6632 &  0.7633 & -4.589\\
10^5 & 5.2336 &  5.2297 &  5.7226 &  5.7241 &  1.4478 &  1.4104 &  4.2194 &  4.2208 &  0.7909 & -1.131\\
10^6 & 8.2869 &  8.2885 &  9.0725 &  9.0720 &  2.2474 &  2.2354 &  6.6891 &  6.6896 &  0.7981 & -0.239\\
10^7 & 13.1356 & 13.1364 & 14.3785 & 14.3782 &  3.5464 &  3.5429 & 10.6021 & 10.6022 &  0.7996 & -0.046
\end{tabular}
\end{ruledtabular}
\end{table*}

\begin{figure}[htb]
\includegraphics[width=0.965\columnwidth,keepaspectratio]{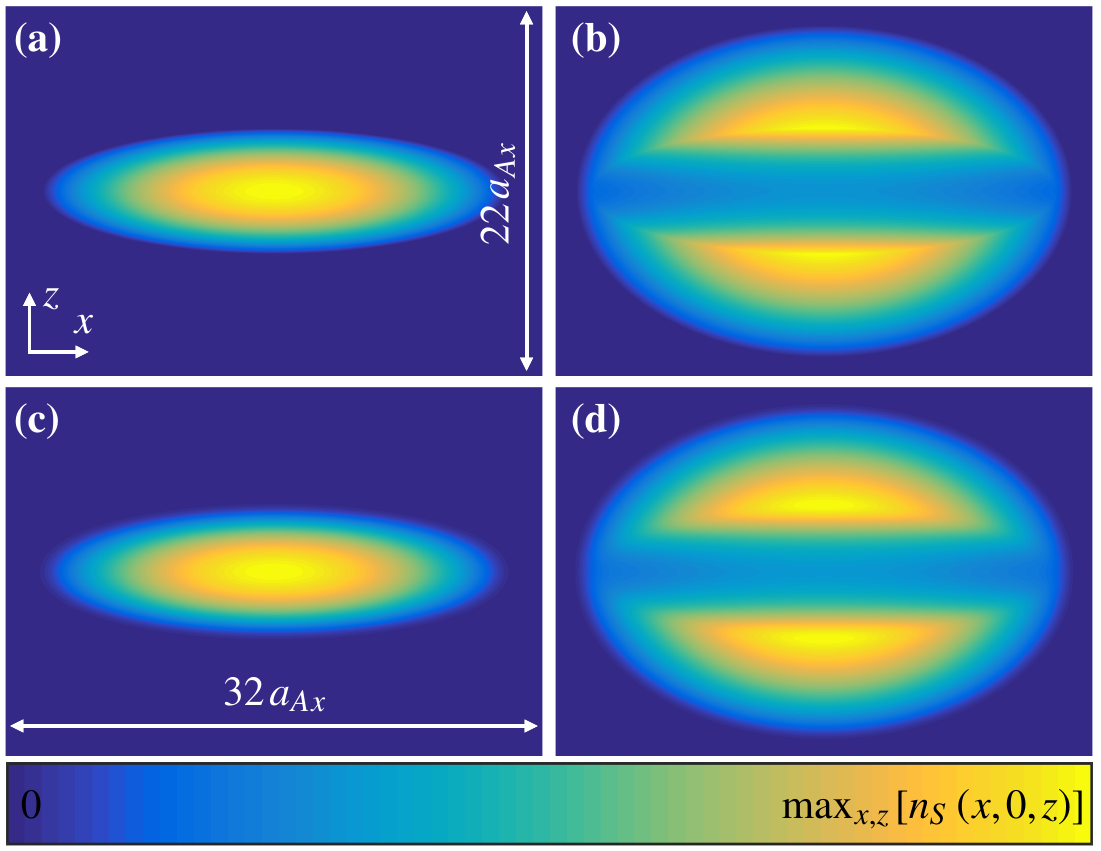}
\caption{\label{fig:oblate-oblate}(a)--(b)~Analytical TF and (c)--(d)~numerical GP solutions for the axisymmetric two-species BEC corresponding to the sixth entry in Table~\ref{table:oblate-oblate}, with $\intra{A} = 10^5$. Panels (a) and (c) show $n_{A}=\abs{\wf_{A}}^2$, while panels (b) and (d) are for $n_{B}=\abs{\wf_{B}}^2$. Here $a_{Ax}=\sqrt{\hbar/\left(m_A\omega_{Ax}\right)}$. Each atomic density is rotationally symmetric about the $z$ axis and is presented here in the plane $y=0$. Evaluating the right-hand side of Eq.~\eqref{ratio2} for the GP solution shown in panels (c) and (d) yields the estimate $\estimate=0.7907$, which is 1.13\% smaller than the true value 0.8.}
\end{figure}

\headerstyle{Interaction strength formula}It turns out that information on $V_{AB}$ can be extracted simply by observing the inner cloud, i.e., the density distribution of species ${A}$. The mean square value of the atomic coordinate $l\in\left\{x,y,z\right\}$ in the species-${A}$ cloud is
\begin{equation}
\secondmoment{A}{l} \equiv \int \abs{\wf _{A}^2} l^{2}\intmeasure =\left(\frac{15}{8\pi }\right)^{2/5}\frac{\formIIgrad{A}{x}^{1/5}\formIIgrad{A}{y}^{1/5}\formIIgrad{A}{z}^{1/5}}{7\,\formIIgrad{A}{l}}.
\label{rms1}
\end{equation}
Furthermore, we define a 3-by-3 matrix $\momentratio{A}{}{}$ with elements
\begin{equation}
\momentratio{A}{l}{l^{\prime}}\equiv \secondmoment{A}{l}/\secondmoment{A}{l^{\prime}}=\formIIgrad{A}{l^{\prime}}/\formIIgrad{A}{l},  \label{d}
\end{equation}
where $l,l^{\prime}\in\left\{x,y,z\right\}$. For the case $\gamma_{Al'}/\gamma_{Al}\neq \gamma_{Bl'}/\gamma_{Bl}$, we can, by using Eqs.~\eqref{eq:formIIgrad}, rewrite Eq.~\eqref{d} as
\begin{equation}
\frac{c_{AB}}{c_{B}}=\frac{m_{A}}{m_{B}}\: \frac{\omega _{Al^{\prime }}^{2}-\omega
_{Al}^{2}\momentratio{A}{l}{l^{\prime}}}{\omega _{Bl^{\prime }}^{2}-\omega
_{Bl}^{2}\momentratio{A}{l}{l^{\prime}}} \qquad \left(\frac{\omega_{Al'}}{\omega_{Al}}\neq \frac{\omega_{Bl'}}{\omega_{Bl}}\right). \label{ratio2}
\end{equation}
If the atomic masses and trap frequencies are known, Eq.~\eqref{ratio2} can be used to determine the relative strength of the interspecies interaction by measuring the mean-square values of any two coordinates in the density distribution of species ${A}$ only. As such, Eq.~\eqref{ratio2} provides a means to determine $c_{AB}$ that complements the standard methods involving collisional measurements on thermal samples~\cite{Fer2002.PRL89.053202,And2005.PRA71.061401}.

We note that the masses and trap frequencies are typically known to a high accuracy (for example, by measuring the dipole oscillations of the center of mass of the atomic cloud, the relative uncertainty in determining the trap frequency can be as small as $0.001$). Consequently, the uncertainties and possible errors in determining $c_{AB}/c_{B}$ via Eq.~\eqref{ratio2} are likely to arise mainly from the uncertainties in the measurement of $\momentratio{A}{l}{l'}$. In the experiments, it is typically straighforward to vary any of the involved trap frequencies and, in this way, to obtain a large number of individual estimates for $c_{AB}/c_{B}$ at different values of the frequencies. Such data will allow one to assess the consistency between Eq.~\eqref{ratio2} and the experimental data and to obtain an accurate estimate for $c_{AB}/c_{B}$ with a well-defined uncertainty by averaging over the individual measurements.

\headerstyle{Numerical results}The above formulae are generalizations to triaxial configurations of previously obtained expressions for spherically symmetric harmonic traps~\cite{Li2017.JPhysB50.135301,He2017.CommunTheorPhys68.220}. They are all based on the TFA and will therefore inherit its errors. However, the TFA is known to become more accurate with increasing number of atoms. It is therefore natural to ask how large condensates one would need in order for the approximation error to be negligible. To this end, we perform numerical calculations beyond the TFA to obtain the exact ground-state solutions of the GP equations~\eqref{eq:GP}. We further define $\estimate$ as the \emph{estimate} obtained from Eq.~\eqref{ratio2} by replacing the TF value of  $\momentratio{A}{l}{l^{\prime}}$ with that of the numerical solution. When the relative error of this estimate is negligible, Eq.~\eqref{ratio2} is applicable for the determination of $c_{AB}/c_{B}$.

In our numerical calculations, we discretize Eqs.~\eqref{eq:GP} by applying the standard three-point finite-difference stencil and solve the resulting equations iteratively with the successive overrelaxation algorithm. We use coordinate grids with step lengths $\leq 0.05 \times \sqrt{\hbar/m_{A}\left(\omega_{Ax}\right)}$ in each direction. To enable simple visualization, we set $\omega_{Ax}=\omega_{Ay}$ and $\omega_{Bx}=\omega_{By}$ and limit the simulations to cases where both $\wf_{A}^2$ and $\wf_{B}^2$ are cylindrically symmetric about the $z$ axis. We stress, however, that our analytical treatment also applies to two-species BECs with no cylindrical symmetry.

Table~\ref{table:oblate-oblate} collects our numerical results for a two-species BEC where the two condensates are coupled through a repulsive interspecies interaction of relative strength $c_{AB}/c_{B}=0.8$ and confined in cylindrically symmetric oblate harmonic traps (i.e., $\omega_{Sz} > \omega_{Sx}=\omega_{Sy}$). The table entries correspond to different values of $\intra{A} \propto N_{A}$, while the other system parameters are kept constant as described in the caption of Table~\ref{table:oblate-oblate}. The analytical TF and the numerical GP solutions for the entry with $\intra{A} = 10^5$ are shown in Fig.~\ref{fig:oblate-oblate}. For the smallest four values of $\intra{A}$ in Table~\ref{table:oblate-oblate}, the root-mean-square (RMS) values $\sqrt{\secondmoment{A}{x}}$ and  $\sqrt{\secondmoment{A}{z}}$ show a noticeable discrepancy between the numerical GP solution and the TFA; consequently, for these states $\estimate$ differs significantly from the true value $0.8$. However, when $\intra{A}$ increases above $10^4$, the accuracy of the TFA improves, the values of $\sqrt{\secondmoment{A}{x}}$ and  $\sqrt{\secondmoment{A}{z}}$ computed for the numerical solution approach their TF limits, and $\estimate$ becomes very close to $0.8$. For $\intra{A} = 10^6$, for instance, $\estimate$ has a relative error of $-0.239\%$ only.

\begin{table*}[t]
\caption{\label{table:prolate-prolate}RMS values of the coordinates $x$ and $z$ in the ground-state density distribution and the corresponding estimates $\estimate$ for a harmonically trapped, three-dimensional two-species BEC with $N_{B}/N_{A}=20$, $c_{AB}/c_{A}=0.8217$, $\omega_{Bx}/\omega_{Ax}=0.65$, $\omega_{Az}/\omega_{Ax}=0.8$, $\omega_{Bz}/\omega_{Bx}=0.6$, and $\omega_{Ay}/\omega_{Ax}=\omega_{By}/\omega_{Bx}=1$. We have used the values $m_{A}/m_{B}=0.471$ and $c_{A}/c_{B}=1.29$, which correspond to species $A$ being  \textsuperscript{41}K and species $B$ being \textsuperscript{87}Rb. The first column shows the value of the intraspecies interaction strength for species $A$, $\intra{A} = N_{A} c_{A}/(\hbar\omega \osclength^3)=4\pi N_{A} a_{AA}/\osclength$, where $\osclength=\sqrt{\hbar/\left(m\omega\right)}$, $m=m_{A}$, and $\omega=\omega_{Ax}$. (If we use $\omega=2\pi\times100$ Hz and $a_{AA}=60\,a_\mathrm{B}$, we obtain $\osclength\approx 1.57$~$\mu\textrm{m}$ and $\intra{A} \approx 0.0254 \times N_{A}$.) All lengths are given in units of $\osclength$, and the true value for $c_{AB}/c_{B}$ is exactly $1.06$.}
\begin{ruledtabular}
\begin{tabular}{ddddddddddd}
 & \multicolumn{2}{c}{~~~~~~~~$\secondmoment{A}{x}^{1/2}$} &  \multicolumn{2}{c}{~~~~~~~~$\secondmoment{B}{x}^{1/2}$} &  \multicolumn{2}{c}{~~~~~~~~$\secondmoment{A}{z}^{1/2}$}  &  \multicolumn{2}{c}{~~~~~~~~$\secondmoment{B}{z}^{1/2}$} & \multicolumn{2}{c}{~~~~$\estimate$}\\
\cline{2-3} \cline{4-5} \cline{6-7} \cline{8-9} \cline{10-11}
\multicolumn{1}{l}{$\intra{A}$} & \multicolumn{1}{r}{Numer.} &  \multicolumn{1}{r}{TFA} &   \multicolumn{1}{r}{Numer.} &  \multicolumn{1}{r}{TFA} &  \multicolumn{1}{r}{Numer.} &  \multicolumn{1}{r}{TFA} &  \multicolumn{1}{r}{Numer.} &  \multicolumn{1}{r}{TFA} & \multicolumn{1}{r}{Value} &  \multicolumn{1}{r}{Error ($\%$)} \\
\hline
10^1 & 1.1053 &  0.9013 &  1.0411 &  1.0061 &  0.9695 &  0.3662 &  1.6973 &  1.7100 &  0.7827 &-26.163\\
10^2 & 1.5755 &  1.4284 &  1.6021 &  1.5946 &  1.0151 &  0.5804 &  2.7023 &  2.7102 &  0.9624 & -9.203\\
10^3 & 2.3155 &  2.2639 &  2.5292 &  2.5272 &  1.1663 &  0.9199 &  4.2918 &  4.2954 &  1.0276 & -3.053\\
10^4 & 3.5987 &  3.5880 &  4.0062 &  4.0054 &  1.5629 &  1.4579 &  6.8064 &  6.8078 &  1.0516 & -0.789\\
10^5 & 5.6867 &  5.6867 &  6.3485 &  6.3481 &  2.3478 &  2.3107 & 10.7891 & 10.7896 &  1.0581 & -0.178\\
10^6 & 9.0117 &  9.0128 & 10.0611 & 10.0610 &  3.6742 &  3.6622 & 17.1002 & 17.1004 &  1.0596 & -0.037\\
10^7 & 14.2838 & 14.2843 & 15.9456 & 15.9456 &  5.8078 &  5.8041 & 27.1022 & 27.1022 & 1.0599 & -0.007
\end{tabular}
\end{ruledtabular}
\end{table*}

\begin{figure}[htb]
\includegraphics[width=0.965\columnwidth,keepaspectratio]{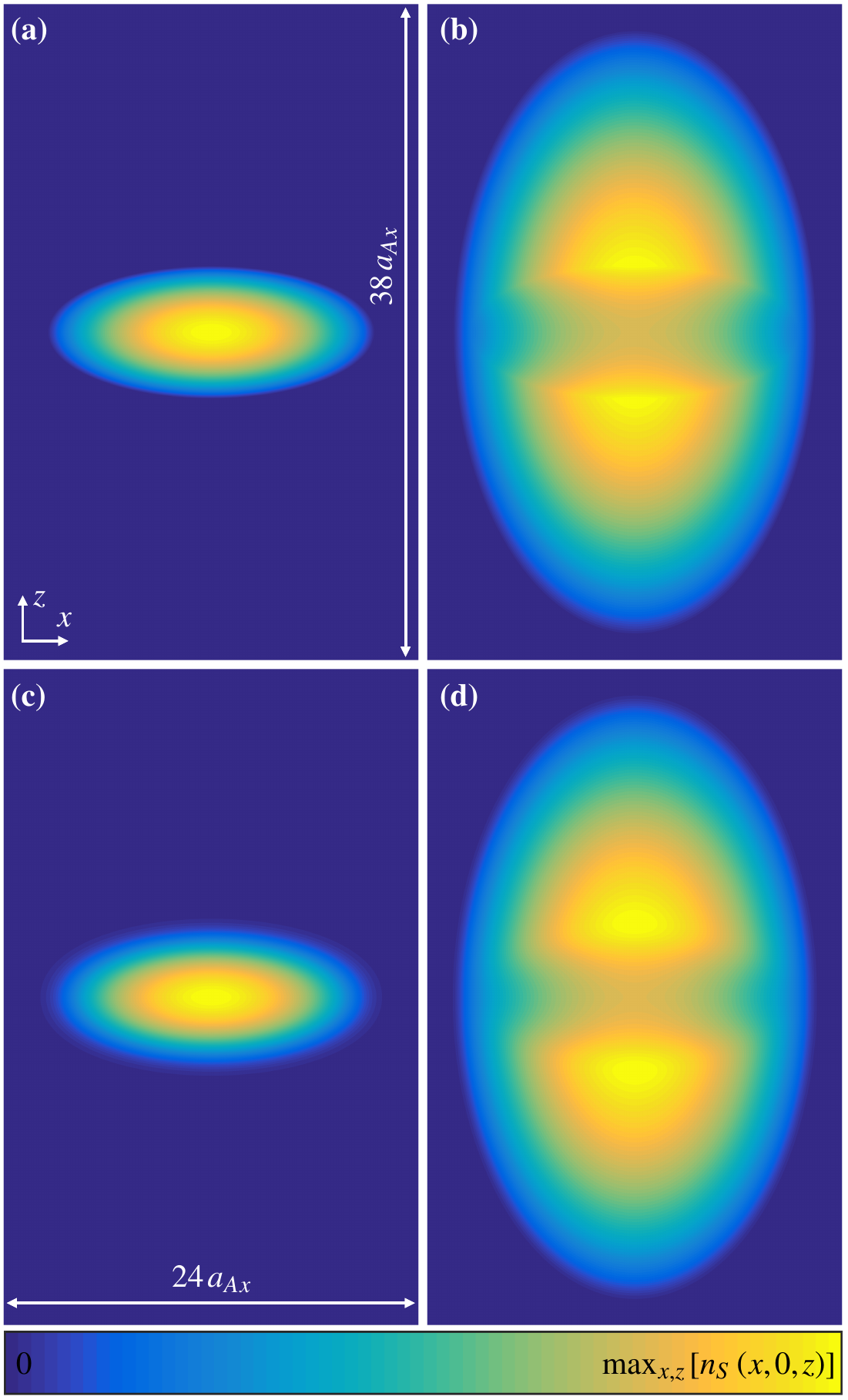}
\caption{\label{fig:prolate-prolate} As Fig.~\ref{fig:oblate-oblate}, but for the fourth entry in Table~\ref{table:prolate-prolate}, with $\intra{A} = 10^4$. Applying Eq.~\eqref{ratio2} to the numerically obtained GP solution in panels (c) and (d) yields the estimate $\estimate=1.0516$, which is 0.79\% smaller than the exact value 1.06.}
\end{figure}

Table~\ref{table:prolate-prolate} lists the corresponding results for a case where both harmonic traps are prolate ($\omega_{Sz} < \omega_{Sx}$) and there is a fairly strong interspecies repulsion of $c_{AB}/c_{B}=1.06$~\footnote{Note that the system satisfies $c_{AB}^2<c_{A}c_{B}$ and is in the miscible regime.}. The analytical and numerical solutions are presented in Fig.~\ref{fig:prolate-prolate} for $\intra{A} = 10^4$. Despite the prolate trap with $\omega_{Az}/\omega_{Ax}=0.8$, the density $\wf_{A}^2$ shown in Fig.~\ref{fig:prolate-prolate}(c) is observed to have a highly oblate profile due to its coupling to species $B$; this suggests that the shape of the cloud carries a strong signal of the interspecies interaction, and consequently we may expect $\estimate$ to be particularly accurate in this configuration. Indeed, its relative error $\estimate\,c_{B}/c_{AB}-1$ is only $-1.13\%$ already at $\intra{A} = 10^4$ and becomes $<10^{-4}$ at $\intra{A} = 10^7$.

\headerstyle{Conclusions}In summary, we have presented an analytical TF solution for a miscible two-species BEC confined in a three-dimensional harmonic trap; we derived a formula, given by Eq.~\eqref{ratio2}, that enables one to determine the relative interspecies strength $c_{AB}/c_{B}$ from the knowledge of the RMS values of two coordinates in the density distribution of species-$A$ atoms. Since Eq.~\eqref{ratio2} holds only within the TFA, we tested its applicability to the numerical solution of the original GP equations. Although the resulting estimate $\estimate$ was found to be highly inaccurate for small numbers of condensed atoms, its relative error became smaller or comparable to typical experimental uncertainties at atom numbers achievable at state-of-the-art experiments. Hence, Eq.~\eqref{ratio2} may provide a useful way of determinining the interspecies interaction strength, complementary to the usual methods that do not necessarily involve BECs. In particular, it could also be used to cross-check that the interspecies interactions between \emph{condensed} atoms are similar to those between noncondensed (but still cold) atoms. 

The basic approach of forming piecewise-defined TF solutions of multispecies BECs is obviously quite general~\cite{Ho1996.PRL77.3276} and can be applied to many more situations besides the one considered here. In future work, it would be beneficial to derive the detailed TF solutions for the entire parameter space and investigate whether similar convenient relations could be found outside the validity range of the present solution. Such a general solution would also facilitate a detailed study of the ground-state phase diagram of the system. Another possible future extension could also be to apply the present approach to mixtures of two \emph{spinor} condensates~\cite{Luo2007.PRA75.043609,Luo2008.JPhysB41.245301,Xu2009.PRA79.023613,Shi2011.PRA83.013616} instead of scalar ones, or to consider a more general scenario where the two harmonic traps are not required to have pairwise parallel symmetry axes.

\begin{acknowledgments}
Useful comments from Dr. Zheng Wei are highly appreciated. This work is supported by the Technology Industries of Finland Centennial Foundation; the Academy of Finland under Grant No.~308632; the National Natural Science Foundation of China under Grants No.~11372122, 11274393, 11574404, and 11275279; the Open Project Program of State Key Laboratory of Theoretical Physics, Institute of Theoretical Physics, Chinese Academy of Sciences, China; and the National Basic Research Program of China (2013CB933601).
\end{acknowledgments}
\bibliographystyle{apsrev4-1}
\bibliography{tc-3d-tfa}

\begin{thebibliography}{54}%
\makeatletter
\providecommand \@ifxundefined [1]{%
 \@ifx{#1\undefined}
}%
\providecommand \@ifnum [1]{%
 \ifnum #1\expandafter \@firstoftwo
 \else \expandafter \@secondoftwo
 \fi
}%
\providecommand \@ifx [1]{%
 \ifx #1\expandafter \@firstoftwo
 \else \expandafter \@secondoftwo
 \fi
}%
\providecommand \natexlab [1]{#1}%
\providecommand \enquote  [1]{``#1''}%
\providecommand \bibnamefont  [1]{#1}%
\providecommand \bibfnamefont [1]{#1}%
\providecommand \citenamefont [1]{#1}%
\providecommand \href@noop [0]{\@secondoftwo}%
\providecommand \href [0]{\begingroup \@sanitize@url \@href}%
\providecommand \@href[1]{\@@startlink{#1}\@@href}%
\providecommand \@@href[1]{\endgroup#1\@@endlink}%
\providecommand \@sanitize@url [0]{\catcode `\\12\catcode `\$12\catcode
  `\&12\catcode `\#12\catcode `\^12\catcode `\_12\catcode `\%12\relax}%
\providecommand \@@startlink[1]{}%
\providecommand \@@endlink[0]{}%
\providecommand \url  [0]{\begingroup\@sanitize@url \@url }%
\providecommand \@url [1]{\endgroup\@href {#1}{\urlprefix }}%
\providecommand \urlprefix  [0]{URL }%
\providecommand \Eprint [0]{\href }%
\providecommand \doibase [0]{http://dx.doi.org/}%
\providecommand \selectlanguage [0]{\@gobble}%
\providecommand \bibinfo  [0]{\@secondoftwo}%
\providecommand \bibfield  [0]{\@secondoftwo}%
\providecommand \translation [1]{[#1]}%
\providecommand \BibitemOpen [0]{}%
\providecommand \bibitemStop [0]{}%
\providecommand \bibitemNoStop [0]{.\EOS\space}%
\providecommand \EOS [0]{\spacefactor3000\relax}%
\providecommand \BibitemShut  [1]{\csname bibitem#1\endcsname}%
\let\auto@bib@innerbib\@empty
\bibitem [{\citenamefont {Ferrari}\ \emph {et~al.}(2002)\citenamefont
  {Ferrari}, \citenamefont {Inguscio}, \citenamefont {Jastrzebski},
  \citenamefont {Modugno}, \citenamefont {Roati},\ and\ \citenamefont
  {Simoni}}]{Fer2002.PRL89.053202}%
  \BibitemOpen
  \bibfield  {author} {\bibinfo {author} {\bibfnamefont {G.}~\bibnamefont
  {Ferrari}}, \bibinfo {author} {\bibfnamefont {M.}~\bibnamefont {Inguscio}},
  \bibinfo {author} {\bibfnamefont {W.}~\bibnamefont {Jastrzebski}}, \bibinfo
  {author} {\bibfnamefont {G.}~\bibnamefont {Modugno}}, \bibinfo {author}
  {\bibfnamefont {G.}~\bibnamefont {Roati}}, \ and\ \bibinfo {author}
  {\bibfnamefont {A.}~\bibnamefont {Simoni}},\ }\href {\doibase
  10.1103/PhysRevLett.89.053202} {\bibfield  {journal} {\bibinfo  {journal}
  {Phys. Rev. Lett.}\ }\textbf {\bibinfo {volume} {89}},\ \bibinfo {pages}
  {053202} (\bibinfo {year} {2002})}\BibitemShut {NoStop}%
\bibitem [{\citenamefont {Modugno}\ \emph {et~al.}(2002)\citenamefont
  {Modugno}, \citenamefont {Modugno}, \citenamefont {Riboli}, \citenamefont
  {Roati},\ and\ \citenamefont {Inguscio}}]{Mod2002.PRL89.190404}%
  \BibitemOpen
  \bibfield  {author} {\bibinfo {author} {\bibfnamefont {G.}~\bibnamefont
  {Modugno}}, \bibinfo {author} {\bibfnamefont {M.}~\bibnamefont {Modugno}},
  \bibinfo {author} {\bibfnamefont {F.}~\bibnamefont {Riboli}}, \bibinfo
  {author} {\bibfnamefont {G.}~\bibnamefont {Roati}}, \ and\ \bibinfo {author}
  {\bibfnamefont {M.}~\bibnamefont {Inguscio}},\ }\href {\doibase
  10.1103/PhysRevLett.89.190404} {\bibfield  {journal} {\bibinfo  {journal}
  {Phys. Rev. Lett.}\ }\textbf {\bibinfo {volume} {89}},\ \bibinfo {pages}
  {190404} (\bibinfo {year} {2002})}\BibitemShut {NoStop}%
\bibitem [{\citenamefont {Thalhammer}\ \emph {et~al.}(2008)\citenamefont
  {Thalhammer}, \citenamefont {Barontini}, \citenamefont {De~Sarlo},
  \citenamefont {Catani}, \citenamefont {Minardi},\ and\ \citenamefont
  {Inguscio}}]{Tha2008.PRL100.210402}%
  \BibitemOpen
  \bibfield  {author} {\bibinfo {author} {\bibfnamefont {G.}~\bibnamefont
  {Thalhammer}}, \bibinfo {author} {\bibfnamefont {G.}~\bibnamefont
  {Barontini}}, \bibinfo {author} {\bibfnamefont {L.}~\bibnamefont {De~Sarlo}},
  \bibinfo {author} {\bibfnamefont {J.}~\bibnamefont {Catani}}, \bibinfo
  {author} {\bibfnamefont {F.}~\bibnamefont {Minardi}}, \ and\ \bibinfo
  {author} {\bibfnamefont {M.}~\bibnamefont {Inguscio}},\ }\href {\doibase
  10.1103/PhysRevLett.100.210402} {\bibfield  {journal} {\bibinfo  {journal}
  {Phys. Rev. Lett.}\ }\textbf {\bibinfo {volume} {100}},\ \bibinfo {pages}
  {210402} (\bibinfo {year} {2008})}\BibitemShut {NoStop}%
\bibitem [{\citenamefont {Aikawa}\ \emph {et~al.}(2009)\citenamefont {Aikawa},
  \citenamefont {Akamatsu}, \citenamefont {Kobayashi}, \citenamefont {Ueda},
  \citenamefont {Kishimoto},\ and\ \citenamefont
  {Inouye}}]{Aik2009.NJP11.055035}%
  \BibitemOpen
  \bibfield  {author} {\bibinfo {author} {\bibfnamefont {K.}~\bibnamefont
  {Aikawa}}, \bibinfo {author} {\bibfnamefont {D.}~\bibnamefont {Akamatsu}},
  \bibinfo {author} {\bibfnamefont {J.}~\bibnamefont {Kobayashi}}, \bibinfo
  {author} {\bibfnamefont {M.}~\bibnamefont {Ueda}}, \bibinfo {author}
  {\bibfnamefont {T.}~\bibnamefont {Kishimoto}}, \ and\ \bibinfo {author}
  {\bibfnamefont {S.}~\bibnamefont {Inouye}},\ }\href {\doibase
  10.1088/1367-2630/11/5/055035} {\bibfield  {journal} {\bibinfo  {journal}
  {New J. Phys.}\ }\textbf {\bibinfo {volume} {11}},\ \bibinfo {pages} {055035}
  (\bibinfo {year} {2009})}\BibitemShut {NoStop}%
\bibitem [{\citenamefont {Catani}\ \emph {et~al.}(2009)\citenamefont {Catani},
  \citenamefont {Barontini}, \citenamefont {Lamporesi}, \citenamefont
  {Rabatti}, \citenamefont {Thalhammer}, \citenamefont {Minardi}, \citenamefont
  {Stringari},\ and\ \citenamefont {Inguscio}}]{Cat2009.PRL103.140401}%
  \BibitemOpen
  \bibfield  {author} {\bibinfo {author} {\bibfnamefont {J.}~\bibnamefont
  {Catani}}, \bibinfo {author} {\bibfnamefont {G.}~\bibnamefont {Barontini}},
  \bibinfo {author} {\bibfnamefont {G.}~\bibnamefont {Lamporesi}}, \bibinfo
  {author} {\bibfnamefont {F.}~\bibnamefont {Rabatti}}, \bibinfo {author}
  {\bibfnamefont {G.}~\bibnamefont {Thalhammer}}, \bibinfo {author}
  {\bibfnamefont {F.}~\bibnamefont {Minardi}}, \bibinfo {author} {\bibfnamefont
  {S.}~\bibnamefont {Stringari}}, \ and\ \bibinfo {author} {\bibfnamefont
  {M.}~\bibnamefont {Inguscio}},\ }\href {\doibase
  10.1103/PhysRevLett.103.140401} {\bibfield  {journal} {\bibinfo  {journal}
  {Phys. Rev. Lett.}\ }\textbf {\bibinfo {volume} {103}},\ \bibinfo {pages}
  {140401} (\bibinfo {year} {2009})}\BibitemShut {NoStop}%
\bibitem [{\citenamefont {McCarron}\ \emph {et~al.}(2011)\citenamefont
  {McCarron}, \citenamefont {Cho}, \citenamefont {Jenkin}, \citenamefont
  {K\"oppinger},\ and\ \citenamefont {Cornish}}]{Car2011.PRA84.011603}%
  \BibitemOpen
  \bibfield  {author} {\bibinfo {author} {\bibfnamefont {D.~J.}\ \bibnamefont
  {McCarron}}, \bibinfo {author} {\bibfnamefont {H.~W.}\ \bibnamefont {Cho}},
  \bibinfo {author} {\bibfnamefont {D.~L.}\ \bibnamefont {Jenkin}}, \bibinfo
  {author} {\bibfnamefont {M.~P.}\ \bibnamefont {K\"oppinger}}, \ and\ \bibinfo
  {author} {\bibfnamefont {S.~L.}\ \bibnamefont {Cornish}},\ }\href {\doibase
  10.1103/PhysRevA.84.011603} {\bibfield  {journal} {\bibinfo  {journal} {Phys.
  Rev. A}\ }\textbf {\bibinfo {volume} {84}},\ \bibinfo {pages} {011603}
  (\bibinfo {year} {2011})}\BibitemShut {NoStop}%
\bibitem [{\citenamefont {Lercher}\ \emph {et~al.}(2011)\citenamefont
  {Lercher}, \citenamefont {Takekoshi}, \citenamefont {Debatin}, \citenamefont
  {Schuster}, \citenamefont {Rameshan}, \citenamefont {Ferlaino}, \citenamefont
  {Grimm},\ and\ \citenamefont {N\"agerl}}]{Ler2011.EPJD65.3}%
  \BibitemOpen
  \bibfield  {author} {\bibinfo {author} {\bibfnamefont {A.}~\bibnamefont
  {Lercher}}, \bibinfo {author} {\bibfnamefont {T.}~\bibnamefont {Takekoshi}},
  \bibinfo {author} {\bibfnamefont {M.}~\bibnamefont {Debatin}}, \bibinfo
  {author} {\bibfnamefont {B.}~\bibnamefont {Schuster}}, \bibinfo {author}
  {\bibfnamefont {R.}~\bibnamefont {Rameshan}}, \bibinfo {author}
  {\bibfnamefont {F.}~\bibnamefont {Ferlaino}}, \bibinfo {author}
  {\bibfnamefont {R.}~\bibnamefont {Grimm}}, \ and\ \bibinfo {author}
  {\bibfnamefont {H.-C.}\ \bibnamefont {N\"agerl}},\ }\href {\doibase
  10.1140/epjd/e2011-20015-6} {\bibfield  {journal} {\bibinfo  {journal} {Eur.
  Phys. J. D}\ }\textbf {\bibinfo {volume} {65}},\ \bibinfo {pages} {3}
  (\bibinfo {year} {2011})}\BibitemShut {NoStop}%
\bibitem [{\citenamefont {Pasquiou}\ \emph {et~al.}(2013)\citenamefont
  {Pasquiou}, \citenamefont {Bayerle}, \citenamefont {Tzanova}, \citenamefont
  {Stellmer}, \citenamefont {Szczepkowski}, \citenamefont {Parigger},
  \citenamefont {Grimm},\ and\ \citenamefont {Schreck}}]{Pas2013.PRA88.023601}%
  \BibitemOpen
  \bibfield  {author} {\bibinfo {author} {\bibfnamefont {B.}~\bibnamefont
  {Pasquiou}}, \bibinfo {author} {\bibfnamefont {A.}~\bibnamefont {Bayerle}},
  \bibinfo {author} {\bibfnamefont {S.~M.}\ \bibnamefont {Tzanova}}, \bibinfo
  {author} {\bibfnamefont {S.}~\bibnamefont {Stellmer}}, \bibinfo {author}
  {\bibfnamefont {J.}~\bibnamefont {Szczepkowski}}, \bibinfo {author}
  {\bibfnamefont {M.}~\bibnamefont {Parigger}}, \bibinfo {author}
  {\bibfnamefont {R.}~\bibnamefont {Grimm}}, \ and\ \bibinfo {author}
  {\bibfnamefont {F.}~\bibnamefont {Schreck}},\ }\href {\doibase
  10.1103/PhysRevA.88.023601} {\bibfield  {journal} {\bibinfo  {journal} {Phys.
  Rev. A}\ }\textbf {\bibinfo {volume} {88}},\ \bibinfo {pages} {023601}
  (\bibinfo {year} {2013})}\BibitemShut {NoStop}%
\bibitem [{\citenamefont {Wacker}\ \emph {et~al.}(2015)\citenamefont {Wacker},
  \citenamefont {J\o{}rgensen}, \citenamefont {Birkmose}, \citenamefont
  {Horchani}, \citenamefont {Ertmer}, \citenamefont {Klempt}, \citenamefont
  {Winter}, \citenamefont {Sherson},\ and\ \citenamefont
  {Arlt}}]{Wac2015.PRA92.053602}%
  \BibitemOpen
  \bibfield  {author} {\bibinfo {author} {\bibfnamefont {L.}~\bibnamefont
  {Wacker}}, \bibinfo {author} {\bibfnamefont {N.~B.}\ \bibnamefont
  {J\o{}rgensen}}, \bibinfo {author} {\bibfnamefont {D.}~\bibnamefont
  {Birkmose}}, \bibinfo {author} {\bibfnamefont {R.}~\bibnamefont {Horchani}},
  \bibinfo {author} {\bibfnamefont {W.}~\bibnamefont {Ertmer}}, \bibinfo
  {author} {\bibfnamefont {C.}~\bibnamefont {Klempt}}, \bibinfo {author}
  {\bibfnamefont {N.}~\bibnamefont {Winter}}, \bibinfo {author} {\bibfnamefont
  {J.}~\bibnamefont {Sherson}}, \ and\ \bibinfo {author} {\bibfnamefont
  {J.~J.}\ \bibnamefont {Arlt}},\ }\href {\doibase 10.1103/PhysRevA.92.053602}
  {\bibfield  {journal} {\bibinfo  {journal} {Phys. Rev. A}\ }\textbf {\bibinfo
  {volume} {92}},\ \bibinfo {pages} {053602} (\bibinfo {year}
  {2015})}\BibitemShut {NoStop}%
\bibitem [{\citenamefont {Wang}\ \emph {et~al.}(2015)\citenamefont {Wang},
  \citenamefont {Li}, \citenamefont {Xiong},\ and\ \citenamefont
  {Wang}}]{Wan2015.JPhysB49.015302}%
  \BibitemOpen
  \bibfield  {author} {\bibinfo {author} {\bibfnamefont {F.}~\bibnamefont
  {Wang}}, \bibinfo {author} {\bibfnamefont {X.}~\bibnamefont {Li}}, \bibinfo
  {author} {\bibfnamefont {D.}~\bibnamefont {Xiong}}, \ and\ \bibinfo {author}
  {\bibfnamefont {D.}~\bibnamefont {Wang}},\ }\href {\doibase
  10.1088/0953-4075/49/1/015302} {\bibfield  {journal} {\bibinfo  {journal} {J.
  Phys. B: At. Mol. Opt. Phys.}\ }\textbf {\bibinfo {volume} {49}},\ \bibinfo
  {pages} {015302} (\bibinfo {year} {2015})}\BibitemShut {NoStop}%
\bibitem [{\citenamefont {Papp}\ \emph {et~al.}(2008)\citenamefont {Papp},
  \citenamefont {Pino},\ and\ \citenamefont {Wieman}}]{Pap2008.PRL101.040402}%
  \BibitemOpen
  \bibfield  {author} {\bibinfo {author} {\bibfnamefont {S.~B.}\ \bibnamefont
  {Papp}}, \bibinfo {author} {\bibfnamefont {J.~M.}\ \bibnamefont {Pino}}, \
  and\ \bibinfo {author} {\bibfnamefont {C.~E.}\ \bibnamefont {Wieman}},\
  }\href {\doibase 10.1103/PhysRevLett.101.040402} {\bibfield  {journal}
  {\bibinfo  {journal} {Phys. Rev. Lett.}\ }\textbf {\bibinfo {volume} {101}},\
  \bibinfo {pages} {040402} (\bibinfo {year} {2008})}\BibitemShut {NoStop}%
\bibitem [{\citenamefont {Sugawa}\ \emph {et~al.}(2011)\citenamefont {Sugawa},
  \citenamefont {Yamazaki}, \citenamefont {Taie},\ and\ \citenamefont
  {Takahashi}}]{Sug2011.PRA84.011610}%
  \BibitemOpen
  \bibfield  {author} {\bibinfo {author} {\bibfnamefont {S.}~\bibnamefont
  {Sugawa}}, \bibinfo {author} {\bibfnamefont {R.}~\bibnamefont {Yamazaki}},
  \bibinfo {author} {\bibfnamefont {S.}~\bibnamefont {Taie}}, \ and\ \bibinfo
  {author} {\bibfnamefont {Y.}~\bibnamefont {Takahashi}},\ }\href {\doibase
  10.1103/PhysRevA.84.011610} {\bibfield  {journal} {\bibinfo  {journal} {Phys.
  Rev. A}\ }\textbf {\bibinfo {volume} {84}},\ \bibinfo {pages} {011610}
  (\bibinfo {year} {2011})}\BibitemShut {NoStop}%
\bibitem [{\citenamefont {Stellmer}\ \emph {et~al.}(2013)\citenamefont
  {Stellmer}, \citenamefont {Grimm},\ and\ \citenamefont
  {Schreck}}]{Ste2013.PRA87.013611}%
  \BibitemOpen
  \bibfield  {author} {\bibinfo {author} {\bibfnamefont {S.}~\bibnamefont
  {Stellmer}}, \bibinfo {author} {\bibfnamefont {R.}~\bibnamefont {Grimm}}, \
  and\ \bibinfo {author} {\bibfnamefont {F.}~\bibnamefont {Schreck}},\ }\href
  {\doibase 10.1103/PhysRevA.87.013611} {\bibfield  {journal} {\bibinfo
  {journal} {Phys. Rev. A}\ }\textbf {\bibinfo {volume} {87}},\ \bibinfo
  {pages} {013611} (\bibinfo {year} {2013})}\BibitemShut {NoStop}%
\bibitem [{\citenamefont {Myatt}\ \emph {et~al.}(1997)\citenamefont {Myatt},
  \citenamefont {Burt}, \citenamefont {Ghrist}, \citenamefont {Cornell},\ and\
  \citenamefont {Wieman}}]{Mya1997.PRL78.586}%
  \BibitemOpen
  \bibfield  {author} {\bibinfo {author} {\bibfnamefont {C.~J.}\ \bibnamefont
  {Myatt}}, \bibinfo {author} {\bibfnamefont {E.~A.}\ \bibnamefont {Burt}},
  \bibinfo {author} {\bibfnamefont {R.~W.}\ \bibnamefont {Ghrist}}, \bibinfo
  {author} {\bibfnamefont {E.~A.}\ \bibnamefont {Cornell}}, \ and\ \bibinfo
  {author} {\bibfnamefont {C.~E.}\ \bibnamefont {Wieman}},\ }\href {\doibase
  10.1103/PhysRevLett.78.586} {\bibfield  {journal} {\bibinfo  {journal} {Phys.
  Rev. Lett.}\ }\textbf {\bibinfo {volume} {78}},\ \bibinfo {pages} {586}
  (\bibinfo {year} {1997})}\BibitemShut {NoStop}%
\bibitem [{\citenamefont {Hall}\ \emph {et~al.}(1998)\citenamefont {Hall},
  \citenamefont {Matthews}, \citenamefont {Ensher}, \citenamefont {Wieman},\
  and\ \citenamefont {Cornell}}]{Hal1998.PRL81.1539}%
  \BibitemOpen
  \bibfield  {author} {\bibinfo {author} {\bibfnamefont {D.~S.}\ \bibnamefont
  {Hall}}, \bibinfo {author} {\bibfnamefont {M.~R.}\ \bibnamefont {Matthews}},
  \bibinfo {author} {\bibfnamefont {J.~R.}\ \bibnamefont {Ensher}}, \bibinfo
  {author} {\bibfnamefont {C.~E.}\ \bibnamefont {Wieman}}, \ and\ \bibinfo
  {author} {\bibfnamefont {E.~A.}\ \bibnamefont {Cornell}},\ }\href {\doibase
  10.1103/PhysRevLett.81.1539} {\bibfield  {journal} {\bibinfo  {journal}
  {Phys. Rev. Lett.}\ }\textbf {\bibinfo {volume} {81}},\ \bibinfo {pages}
  {1539} (\bibinfo {year} {1998})}\BibitemShut {NoStop}%
\bibitem [{\citenamefont {Matthews}\ \emph {et~al.}(1999)\citenamefont
  {Matthews}, \citenamefont {Anderson}, \citenamefont {Haljan}, \citenamefont
  {Hall}, \citenamefont {Wieman},\ and\ \citenamefont
  {Cornell}}]{Mat1999.PRL83.2498}%
  \BibitemOpen
  \bibfield  {author} {\bibinfo {author} {\bibfnamefont {M.~R.}\ \bibnamefont
  {Matthews}}, \bibinfo {author} {\bibfnamefont {B.~P.}\ \bibnamefont
  {Anderson}}, \bibinfo {author} {\bibfnamefont {P.~C.}\ \bibnamefont
  {Haljan}}, \bibinfo {author} {\bibfnamefont {D.~S.}\ \bibnamefont {Hall}},
  \bibinfo {author} {\bibfnamefont {C.~E.}\ \bibnamefont {Wieman}}, \ and\
  \bibinfo {author} {\bibfnamefont {E.~A.}\ \bibnamefont {Cornell}},\ }\href
  {\doibase 10.1103/PhysRevLett.83.2498} {\bibfield  {journal} {\bibinfo
  {journal} {Phys. Rev. Lett.}\ }\textbf {\bibinfo {volume} {83}},\ \bibinfo
  {pages} {2498} (\bibinfo {year} {1999})}\BibitemShut {NoStop}%
\bibitem [{\citenamefont {Delannoy}\ \emph {et~al.}(2001)\citenamefont
  {Delannoy}, \citenamefont {Murdoch}, \citenamefont {Boyer}, \citenamefont
  {Josse}, \citenamefont {Bouyer},\ and\ \citenamefont
  {Aspect}}]{Del2001.PRA63.051602}%
  \BibitemOpen
  \bibfield  {author} {\bibinfo {author} {\bibfnamefont {G.}~\bibnamefont
  {Delannoy}}, \bibinfo {author} {\bibfnamefont {S.~G.}\ \bibnamefont
  {Murdoch}}, \bibinfo {author} {\bibfnamefont {V.}~\bibnamefont {Boyer}},
  \bibinfo {author} {\bibfnamefont {V.}~\bibnamefont {Josse}}, \bibinfo
  {author} {\bibfnamefont {P.}~\bibnamefont {Bouyer}}, \ and\ \bibinfo {author}
  {\bibfnamefont {A.}~\bibnamefont {Aspect}},\ }\href {\doibase
  10.1103/PhysRevA.63.051602} {\bibfield  {journal} {\bibinfo  {journal} {Phys.
  Rev. A}\ }\textbf {\bibinfo {volume} {63}},\ \bibinfo {pages} {051602}
  (\bibinfo {year} {2001})}\BibitemShut {NoStop}%
\bibitem [{\citenamefont {Schweikhard}\ \emph {et~al.}(2004)\citenamefont
  {Schweikhard}, \citenamefont {Coddington}, \citenamefont {Engels},
  \citenamefont {Tung},\ and\ \citenamefont {Cornell}}]{Sch2004.PRL93.210403}%
  \BibitemOpen
  \bibfield  {author} {\bibinfo {author} {\bibfnamefont {V.}~\bibnamefont
  {Schweikhard}}, \bibinfo {author} {\bibfnamefont {I.}~\bibnamefont
  {Coddington}}, \bibinfo {author} {\bibfnamefont {P.}~\bibnamefont {Engels}},
  \bibinfo {author} {\bibfnamefont {S.}~\bibnamefont {Tung}}, \ and\ \bibinfo
  {author} {\bibfnamefont {E.~A.}\ \bibnamefont {Cornell}},\ }\href {\doibase
  10.1103/PhysRevLett.93.210403} {\bibfield  {journal} {\bibinfo  {journal}
  {Phys. Rev. Lett.}\ }\textbf {\bibinfo {volume} {93}},\ \bibinfo {pages}
  {210403} (\bibinfo {year} {2004})}\BibitemShut {NoStop}%
\bibitem [{\citenamefont {Anderson}\ \emph {et~al.}(2009)\citenamefont
  {Anderson}, \citenamefont {Ticknor}, \citenamefont {Sidorov},\ and\
  \citenamefont {Hall}}]{And2009.PRA80.023603}%
  \BibitemOpen
  \bibfield  {author} {\bibinfo {author} {\bibfnamefont {R.~P.}\ \bibnamefont
  {Anderson}}, \bibinfo {author} {\bibfnamefont {C.}~\bibnamefont {Ticknor}},
  \bibinfo {author} {\bibfnamefont {A.~I.}\ \bibnamefont {Sidorov}}, \ and\
  \bibinfo {author} {\bibfnamefont {B.~V.}\ \bibnamefont {Hall}},\ }\href
  {\doibase 10.1103/PhysRevA.80.023603} {\bibfield  {journal} {\bibinfo
  {journal} {Phys. Rev. A}\ }\textbf {\bibinfo {volume} {80}},\ \bibinfo
  {pages} {023603} (\bibinfo {year} {2009})}\BibitemShut {NoStop}%
\bibitem [{\citenamefont {Timmermans}(1998)}]{Tim1998.PRL81.5718}%
  \BibitemOpen
  \bibfield  {author} {\bibinfo {author} {\bibfnamefont {E.}~\bibnamefont
  {Timmermans}},\ }\href {\doibase 10.1103/PhysRevLett.81.5718} {\bibfield
  {journal} {\bibinfo  {journal} {Phys. Rev. Lett.}\ }\textbf {\bibinfo
  {volume} {81}},\ \bibinfo {pages} {5718} (\bibinfo {year}
  {1998})}\BibitemShut {NoStop}%
\bibitem [{\citenamefont {Ao}\ and\ \citenamefont
  {Chui}(1998)}]{Ao1998.PRA58.4836}%
  \BibitemOpen
  \bibfield  {author} {\bibinfo {author} {\bibfnamefont {P.}~\bibnamefont
  {Ao}}\ and\ \bibinfo {author} {\bibfnamefont {S.~T.}\ \bibnamefont {Chui}},\
  }\href {\doibase 10.1103/PhysRevA.58.4836} {\bibfield  {journal} {\bibinfo
  {journal} {Phys. Rev. A}\ }\textbf {\bibinfo {volume} {58}},\ \bibinfo
  {pages} {4836} (\bibinfo {year} {1998})}\BibitemShut {NoStop}%
\bibitem [{\citenamefont {Indekeu}\ \emph {et~al.}(2015)\citenamefont
  {Indekeu}, \citenamefont {Lin}, \citenamefont {Van~Thu}, \citenamefont
  {Van~Schaeybroeck},\ and\ \citenamefont {Phat}}]{Ind2015.PRA91.033615}%
  \BibitemOpen
  \bibfield  {author} {\bibinfo {author} {\bibfnamefont {J.~O.}\ \bibnamefont
  {Indekeu}}, \bibinfo {author} {\bibfnamefont {C.-Y.}\ \bibnamefont {Lin}},
  \bibinfo {author} {\bibfnamefont {N.}~\bibnamefont {Van~Thu}}, \bibinfo
  {author} {\bibfnamefont {B.}~\bibnamefont {Van~Schaeybroeck}}, \ and\
  \bibinfo {author} {\bibfnamefont {T.~H.}\ \bibnamefont {Phat}},\ }\href
  {\doibase 10.1103/PhysRevA.91.033615} {\bibfield  {journal} {\bibinfo
  {journal} {Phys. Rev. A}\ }\textbf {\bibinfo {volume} {91}},\ \bibinfo
  {pages} {033615} (\bibinfo {year} {2015})}\BibitemShut {NoStop}%
\bibitem [{\citenamefont {Roy}\ and\ \citenamefont
  {Angom}(2015)}]{Roy2015.PRA92.011601}%
  \BibitemOpen
  \bibfield  {author} {\bibinfo {author} {\bibfnamefont {A.}~\bibnamefont
  {Roy}}\ and\ \bibinfo {author} {\bibfnamefont {D.}~\bibnamefont {Angom}},\
  }\href {\doibase 10.1103/PhysRevA.92.011601} {\bibfield  {journal} {\bibinfo
  {journal} {Phys. Rev. A}\ }\textbf {\bibinfo {volume} {92}},\ \bibinfo
  {pages} {011601} (\bibinfo {year} {2015})}\BibitemShut {NoStop}%
\bibitem [{\citenamefont {Polo}\ \emph {et~al.}(2015)\citenamefont {Polo},
  \citenamefont {Ahufinger}, \citenamefont {Mason}, \citenamefont {Sridhar},
  \citenamefont {Billam},\ and\ \citenamefont
  {Gardiner}}]{Pol2015.PRA91.053626}%
  \BibitemOpen
  \bibfield  {author} {\bibinfo {author} {\bibfnamefont {J.}~\bibnamefont
  {Polo}}, \bibinfo {author} {\bibfnamefont {V.}~\bibnamefont {Ahufinger}},
  \bibinfo {author} {\bibfnamefont {P.}~\bibnamefont {Mason}}, \bibinfo
  {author} {\bibfnamefont {S.}~\bibnamefont {Sridhar}}, \bibinfo {author}
  {\bibfnamefont {T.~P.}\ \bibnamefont {Billam}}, \ and\ \bibinfo {author}
  {\bibfnamefont {S.~A.}\ \bibnamefont {Gardiner}},\ }\href {\doibase
  10.1103/PhysRevA.91.053626} {\bibfield  {journal} {\bibinfo  {journal} {Phys.
  Rev. A}\ }\textbf {\bibinfo {volume} {91}},\ \bibinfo {pages} {053626}
  (\bibinfo {year} {2015})}\BibitemShut {NoStop}%
\bibitem [{\citenamefont {Chui}\ and\ \citenamefont
  {Ao}(1999)}]{Chu1999.PRA59.1473}%
  \BibitemOpen
  \bibfield  {author} {\bibinfo {author} {\bibfnamefont {S.~T.}\ \bibnamefont
  {Chui}}\ and\ \bibinfo {author} {\bibfnamefont {P.}~\bibnamefont {Ao}},\
  }\href {\doibase 10.1103/PhysRevA.59.1473} {\bibfield  {journal} {\bibinfo
  {journal} {Phys. Rev. A}\ }\textbf {\bibinfo {volume} {59}},\ \bibinfo
  {pages} {1473} (\bibinfo {year} {1999})}\BibitemShut {NoStop}%
\bibitem [{\citenamefont {Trippenbach}\ \emph {et~al.}(2000)\citenamefont
  {Trippenbach}, \citenamefont {G{\'o}ral}, \citenamefont {Rzazewski},
  \citenamefont {Malomed},\ and\ \citenamefont {Band}}]{Tri2000.JPhysB33.4017}%
  \BibitemOpen
  \bibfield  {author} {\bibinfo {author} {\bibfnamefont {M.}~\bibnamefont
  {Trippenbach}}, \bibinfo {author} {\bibfnamefont {K.}~\bibnamefont
  {G{\'o}ral}}, \bibinfo {author} {\bibfnamefont {K.}~\bibnamefont
  {Rzazewski}}, \bibinfo {author} {\bibfnamefont {B.}~\bibnamefont {Malomed}},
  \ and\ \bibinfo {author} {\bibfnamefont {Y.}~\bibnamefont {Band}},\ }\href
  {\doibase 10.1088/0953-4075/33/19/314} {\bibfield  {journal} {\bibinfo
  {journal} {J. Phys. B: At. Mol. Opt. Phys.}\ }\textbf {\bibinfo {volume}
  {33}},\ \bibinfo {pages} {4017} (\bibinfo {year} {2000})}\BibitemShut
  {NoStop}%
\bibitem [{\citenamefont {Riboli}\ and\ \citenamefont
  {Modugno}(2002)}]{Rib2002.PRA65.063614}%
  \BibitemOpen
  \bibfield  {author} {\bibinfo {author} {\bibfnamefont {F.}~\bibnamefont
  {Riboli}}\ and\ \bibinfo {author} {\bibfnamefont {M.}~\bibnamefont
  {Modugno}},\ }\href {\doibase 10.1103/PhysRevA.65.063614} {\bibfield
  {journal} {\bibinfo  {journal} {Phys. Rev. A}\ }\textbf {\bibinfo {volume}
  {65}},\ \bibinfo {pages} {063614} (\bibinfo {year} {2002})}\BibitemShut
  {NoStop}%
\bibitem [{\citenamefont {Svidzinsky}\ and\ \citenamefont
  {Chui}(2003)}]{Svi2003.PRA67.053608}%
  \BibitemOpen
  \bibfield  {author} {\bibinfo {author} {\bibfnamefont {A.~A.}\ \bibnamefont
  {Svidzinsky}}\ and\ \bibinfo {author} {\bibfnamefont {S.~T.}\ \bibnamefont
  {Chui}},\ }\href {\doibase 10.1103/PhysRevA.67.053608} {\bibfield  {journal}
  {\bibinfo  {journal} {Phys. Rev. A}\ }\textbf {\bibinfo {volume} {67}},\
  \bibinfo {pages} {053608} (\bibinfo {year} {2003})}\BibitemShut {NoStop}%
\bibitem [{\citenamefont {Gautam}\ and\ \citenamefont
  {Angom}(2010)}]{Gau2010.JPhysB43.095302}%
  \BibitemOpen
  \bibfield  {author} {\bibinfo {author} {\bibfnamefont {S.}~\bibnamefont
  {Gautam}}\ and\ \bibinfo {author} {\bibfnamefont {D.}~\bibnamefont {Angom}},\
  }\href {\doibase 10.1088/0953-4075/43/9/095302} {\bibfield  {journal}
  {\bibinfo  {journal} {J. Phys. B: At. Mol. Opt. Phys.}\ }\textbf {\bibinfo
  {volume} {43}},\ \bibinfo {pages} {095302} (\bibinfo {year}
  {2010})}\BibitemShut {NoStop}%
\bibitem [{\citenamefont {Van~Schaeybroeck}\ and\ \citenamefont
  {Indekeu}(2015)}]{Sch2015.PRA91.013626}%
  \BibitemOpen
  \bibfield  {author} {\bibinfo {author} {\bibfnamefont {B.}~\bibnamefont
  {Van~Schaeybroeck}}\ and\ \bibinfo {author} {\bibfnamefont {J.~O.}\
  \bibnamefont {Indekeu}},\ }\href {\doibase 10.1103/PhysRevA.91.013626}
  {\bibfield  {journal} {\bibinfo  {journal} {Phys. Rev. A}\ }\textbf {\bibinfo
  {volume} {91}},\ \bibinfo {pages} {013626} (\bibinfo {year}
  {2015})}\BibitemShut {NoStop}%
\bibitem [{\citenamefont {Mueller}\ and\ \citenamefont
  {Ho}(2002)}]{Mue2002.PRL88.180403}%
  \BibitemOpen
  \bibfield  {author} {\bibinfo {author} {\bibfnamefont {E.~J.}\ \bibnamefont
  {Mueller}}\ and\ \bibinfo {author} {\bibfnamefont {T.-L.}\ \bibnamefont
  {Ho}},\ }\href {\doibase 10.1103/PhysRevLett.88.180403} {\bibfield  {journal}
  {\bibinfo  {journal} {Phys. Rev. Lett.}\ }\textbf {\bibinfo {volume} {88}},\
  \bibinfo {pages} {180403} (\bibinfo {year} {2002})}\BibitemShut {NoStop}%
\bibitem [{\citenamefont {Kasamatsu}\ \emph {et~al.}(2004)\citenamefont
  {Kasamatsu}, \citenamefont {Tsubota},\ and\ \citenamefont
  {Ueda}}]{Kas2004.PRL93.250406}%
  \BibitemOpen
  \bibfield  {author} {\bibinfo {author} {\bibfnamefont {K.}~\bibnamefont
  {Kasamatsu}}, \bibinfo {author} {\bibfnamefont {M.}~\bibnamefont {Tsubota}},
  \ and\ \bibinfo {author} {\bibfnamefont {M.}~\bibnamefont {Ueda}},\ }\href
  {\doibase 10.1103/PhysRevLett.93.250406} {\bibfield  {journal} {\bibinfo
  {journal} {Phys. Rev. Lett.}\ }\textbf {\bibinfo {volume} {93}},\ \bibinfo
  {pages} {250406} (\bibinfo {year} {2004})}\BibitemShut {NoStop}%
\bibitem [{\citenamefont {Kasamatsu}\ \emph
  {et~al.}(2005{\natexlab{a}})\citenamefont {Kasamatsu}, \citenamefont
  {Tsubota},\ and\ \citenamefont {Ueda}}]{Kas2005.PRA71.043611}%
  \BibitemOpen
  \bibfield  {author} {\bibinfo {author} {\bibfnamefont {K.}~\bibnamefont
  {Kasamatsu}}, \bibinfo {author} {\bibfnamefont {M.}~\bibnamefont {Tsubota}},
  \ and\ \bibinfo {author} {\bibfnamefont {M.}~\bibnamefont {Ueda}},\ }\href
  {\doibase 10.1103/PhysRevA.71.043611} {\bibfield  {journal} {\bibinfo
  {journal} {Phys. Rev. A}\ }\textbf {\bibinfo {volume} {71}},\ \bibinfo
  {pages} {043611} (\bibinfo {year} {2005}{\natexlab{a}})}\BibitemShut
  {NoStop}%
\bibitem [{\citenamefont {Kasamatsu}\ \emph
  {et~al.}(2005{\natexlab{b}})\citenamefont {Kasamatsu}, \citenamefont
  {Tsubota},\ and\ \citenamefont {Ueda}}]{Kas2005.IJMPB19.1835}%
  \BibitemOpen
  \bibfield  {author} {\bibinfo {author} {\bibfnamefont {K.}~\bibnamefont
  {Kasamatsu}}, \bibinfo {author} {\bibfnamefont {M.}~\bibnamefont {Tsubota}},
  \ and\ \bibinfo {author} {\bibfnamefont {M.}~\bibnamefont {Ueda}},\ }\href
  {\doibase 10.1142/S0217979205029602} {\bibfield  {journal} {\bibinfo
  {journal} {Int. J. Mod. Phys. B}\ }\textbf {\bibinfo {volume} {19}},\
  \bibinfo {pages} {1835} (\bibinfo {year} {2005}{\natexlab{b}})}\BibitemShut
  {NoStop}%
\bibitem [{\citenamefont {Yang}\ \emph {et~al.}(2008)\citenamefont {Yang},
  \citenamefont {Wu}, \citenamefont {Zhang},\ and\ \citenamefont
  {Feng}}]{Yan2008.PRA77.033621}%
  \BibitemOpen
  \bibfield  {author} {\bibinfo {author} {\bibfnamefont {S.-J.}\ \bibnamefont
  {Yang}}, \bibinfo {author} {\bibfnamefont {Q.-S.}\ \bibnamefont {Wu}},
  \bibinfo {author} {\bibfnamefont {S.-N.}\ \bibnamefont {Zhang}}, \ and\
  \bibinfo {author} {\bibfnamefont {S.}~\bibnamefont {Feng}},\ }\href {\doibase
  10.1103/PhysRevA.77.033621} {\bibfield  {journal} {\bibinfo  {journal} {Phys.
  Rev. A}\ }\textbf {\bibinfo {volume} {77}},\ \bibinfo {pages} {033621}
  (\bibinfo {year} {2008})}\BibitemShut {NoStop}%
\bibitem [{\citenamefont {Kasamatsu}\ and\ \citenamefont
  {Tsubota}(2009)}]{Kas2009.PRA79.023606}%
  \BibitemOpen
  \bibfield  {author} {\bibinfo {author} {\bibfnamefont {K.}~\bibnamefont
  {Kasamatsu}}\ and\ \bibinfo {author} {\bibfnamefont {M.}~\bibnamefont
  {Tsubota}},\ }\href {\doibase 10.1103/PhysRevA.79.023606} {\bibfield
  {journal} {\bibinfo  {journal} {Phys. Rev. A}\ }\textbf {\bibinfo {volume}
  {79}},\ \bibinfo {pages} {023606} (\bibinfo {year} {2009})}\BibitemShut
  {NoStop}%
\bibitem [{\citenamefont {Mason}\ and\ \citenamefont
  {Aftalion}(2011)}]{Mas2011.PRA84.033611}%
  \BibitemOpen
  \bibfield  {author} {\bibinfo {author} {\bibfnamefont {P.}~\bibnamefont
  {Mason}}\ and\ \bibinfo {author} {\bibfnamefont {A.}~\bibnamefont
  {Aftalion}},\ }\href {\doibase 10.1103/PhysRevA.84.033611} {\bibfield
  {journal} {\bibinfo  {journal} {Phys. Rev. A}\ }\textbf {\bibinfo {volume}
  {84}},\ \bibinfo {pages} {033611} (\bibinfo {year} {2011})}\BibitemShut
  {NoStop}%
\bibitem [{\citenamefont {Kuopanportti}\ \emph {et~al.}(2012)\citenamefont
  {Kuopanportti}, \citenamefont {Huhtam\"aki},\ and\ \citenamefont
  {M\"ott\"onen}}]{Kuo2012.PRA85.043613}%
  \BibitemOpen
  \bibfield  {author} {\bibinfo {author} {\bibfnamefont {P.}~\bibnamefont
  {Kuopanportti}}, \bibinfo {author} {\bibfnamefont {J.~A.~M.}\ \bibnamefont
  {Huhtam\"aki}}, \ and\ \bibinfo {author} {\bibfnamefont {M.}~\bibnamefont
  {M\"ott\"onen}},\ }\href {\doibase 10.1103/PhysRevA.85.043613} {\bibfield
  {journal} {\bibinfo  {journal} {Phys. Rev. A}\ }\textbf {\bibinfo {volume}
  {85}},\ \bibinfo {pages} {043613} (\bibinfo {year} {2012})}\BibitemShut
  {NoStop}%
\bibitem [{\citenamefont {Kuopanportti}\ \emph {et~al.}(2015)\citenamefont
  {Kuopanportti}, \citenamefont {Orlova},\ and\ \citenamefont {Milo\ifmmode
  \check{s}\else \v{s}\fi{}evi\ifmmode~\acute{c}\else
  \'{c}\fi{}}}]{Kuo2015.PRA91.043605}%
  \BibitemOpen
  \bibfield  {author} {\bibinfo {author} {\bibfnamefont {P.}~\bibnamefont
  {Kuopanportti}}, \bibinfo {author} {\bibfnamefont {N.~V.}\ \bibnamefont
  {Orlova}}, \ and\ \bibinfo {author} {\bibfnamefont {M.~V.}\ \bibnamefont
  {Milo\ifmmode \check{s}\else \v{s}\fi{}evi\ifmmode~\acute{c}\else
  \'{c}\fi{}}},\ }\href {\doibase 10.1103/PhysRevA.91.043605} {\bibfield
  {journal} {\bibinfo  {journal} {Phys. Rev. A}\ }\textbf {\bibinfo {volume}
  {91}},\ \bibinfo {pages} {043605} (\bibinfo {year} {2015})}\BibitemShut
  {NoStop}%
\bibitem [{\citenamefont {Galteland}\ \emph {et~al.}(2015)\citenamefont
  {Galteland}, \citenamefont {Babaev},\ and\ \citenamefont
  {Sudb{\o}}}]{Gal2015.NJP17.103040}%
  \BibitemOpen
  \bibfield  {author} {\bibinfo {author} {\bibfnamefont {P.~N.}\ \bibnamefont
  {Galteland}}, \bibinfo {author} {\bibfnamefont {E.}~\bibnamefont {Babaev}}, \
  and\ \bibinfo {author} {\bibfnamefont {A.}~\bibnamefont {Sudb{\o}}},\ }\href
  {\doibase 10.1088/1367-2630/17/10/103040} {\bibfield  {journal} {\bibinfo
  {journal} {New J. Phys.}\ }\textbf {\bibinfo {volume} {17}},\ \bibinfo
  {pages} {103040} (\bibinfo {year} {2015})}\BibitemShut {NoStop}%
\bibitem [{Note1()}]{Note1}%
  \BibitemOpen
  \bibinfo {note} {If the finite size and inhomogeneity of the system are
  neglected, the miscible regime corresponds to $\delimiter 69640972 c_{AB}
  \delimiter 86418188 < \protect \sqrt {c_{A}c_{B}}$.}\BibitemShut {Stop}%
\bibitem [{\citenamefont {Esry}\ \emph {et~al.}(1997)\citenamefont {Esry},
  \citenamefont {Greene}, \citenamefont {Burke},\ and\ \citenamefont
  {Bohn}}]{Esr1997.PRL78.3594}%
  \BibitemOpen
  \bibfield  {author} {\bibinfo {author} {\bibfnamefont {B.~D.}\ \bibnamefont
  {Esry}}, \bibinfo {author} {\bibfnamefont {C.~H.}\ \bibnamefont {Greene}},
  \bibinfo {author} {\bibfnamefont {J.~P.}\ \bibnamefont {Burke}, \bibfnamefont
  {Jr.}}, \ and\ \bibinfo {author} {\bibfnamefont {J.~L.}\ \bibnamefont
  {Bohn}},\ }\href {\doibase 10.1103/PhysRevLett.78.3594} {\bibfield  {journal}
  {\bibinfo  {journal} {Phys. Rev. Lett.}\ }\textbf {\bibinfo {volume} {78}},\
  \bibinfo {pages} {3594} (\bibinfo {year} {1997})}\BibitemShut {NoStop}%
\bibitem [{\citenamefont {Pu}\ and\ \citenamefont
  {Bigelow}(1998)}]{Pu1998.PRL80.1130}%
  \BibitemOpen
  \bibfield  {author} {\bibinfo {author} {\bibfnamefont {H.}~\bibnamefont
  {Pu}}\ and\ \bibinfo {author} {\bibfnamefont {N.~P.}\ \bibnamefont
  {Bigelow}},\ }\href {\doibase 10.1103/PhysRevLett.80.1130} {\bibfield
  {journal} {\bibinfo  {journal} {Phys. Rev. Lett.}\ }\textbf {\bibinfo
  {volume} {80}},\ \bibinfo {pages} {1130} (\bibinfo {year}
  {1998})}\BibitemShut {NoStop}%
\bibitem [{\citenamefont {Ho}\ and\ \citenamefont
  {Shenoy}(1996)}]{Ho1996.PRL77.3276}%
  \BibitemOpen
  \bibfield  {author} {\bibinfo {author} {\bibfnamefont {T.-L.}\ \bibnamefont
  {Ho}}\ and\ \bibinfo {author} {\bibfnamefont {V.~B.}\ \bibnamefont
  {Shenoy}},\ }\href {\doibase 10.1103/PhysRevLett.77.3276} {\bibfield
  {journal} {\bibinfo  {journal} {Phys. Rev. Lett.}\ }\textbf {\bibinfo
  {volume} {77}},\ \bibinfo {pages} {3276} (\bibinfo {year}
  {1996})}\BibitemShut {NoStop}%
\bibitem [{\citenamefont {Edwards}\ and\ \citenamefont
  {Burnett}(1995)}]{Edw1995.PRA51.1382}%
  \BibitemOpen
  \bibfield  {author} {\bibinfo {author} {\bibfnamefont {M.}~\bibnamefont
  {Edwards}}\ and\ \bibinfo {author} {\bibfnamefont {K.}~\bibnamefont
  {Burnett}},\ }\href {\doibase 10.1103/PhysRevA.51.1382} {\bibfield  {journal}
  {\bibinfo  {journal} {Phys. Rev. A}\ }\textbf {\bibinfo {volume} {51}},\
  \bibinfo {pages} {1382} (\bibinfo {year} {1995})}\BibitemShut {NoStop}%
\bibitem [{\citenamefont {Baym}\ and\ \citenamefont
  {Pethick}(1996)}]{Bay1996.PRL76.6}%
  \BibitemOpen
  \bibfield  {author} {\bibinfo {author} {\bibfnamefont {G.}~\bibnamefont
  {Baym}}\ and\ \bibinfo {author} {\bibfnamefont {C.~J.}\ \bibnamefont
  {Pethick}},\ }\href {\doibase 10.1103/PhysRevLett.76.6} {\bibfield  {journal}
  {\bibinfo  {journal} {Phys. Rev. Lett.}\ }\textbf {\bibinfo {volume} {76}},\
  \bibinfo {pages} {6} (\bibinfo {year} {1996})}\BibitemShut {NoStop}%
\bibitem [{\citenamefont {Anderlini}\ \emph {et~al.}(2005)\citenamefont
  {Anderlini}, \citenamefont {Courtade}, \citenamefont {Cristiani},
  \citenamefont {Cossart}, \citenamefont {Ciampini}, \citenamefont {Sias},
  \citenamefont {Morsch},\ and\ \citenamefont
  {Arimondo}}]{And2005.PRA71.061401}%
  \BibitemOpen
  \bibfield  {author} {\bibinfo {author} {\bibfnamefont {M.}~\bibnamefont
  {Anderlini}}, \bibinfo {author} {\bibfnamefont {E.}~\bibnamefont {Courtade}},
  \bibinfo {author} {\bibfnamefont {M.}~\bibnamefont {Cristiani}}, \bibinfo
  {author} {\bibfnamefont {D.}~\bibnamefont {Cossart}}, \bibinfo {author}
  {\bibfnamefont {D.}~\bibnamefont {Ciampini}}, \bibinfo {author}
  {\bibfnamefont {C.}~\bibnamefont {Sias}}, \bibinfo {author} {\bibfnamefont
  {O.}~\bibnamefont {Morsch}}, \ and\ \bibinfo {author} {\bibfnamefont
  {E.}~\bibnamefont {Arimondo}},\ }\href {\doibase 10.1103/PhysRevA.71.061401}
  {\bibfield  {journal} {\bibinfo  {journal} {Phys. Rev. A}\ }\textbf {\bibinfo
  {volume} {71}},\ \bibinfo {pages} {061401} (\bibinfo {year}
  {2005})}\BibitemShut {NoStop}%
\bibitem [{\citenamefont {Li}\ \emph {et~al.}(2017)\citenamefont {Li},
  \citenamefont {Liu}, \citenamefont {Yao},\ and\ \citenamefont
  {Bao}}]{Li2017.JPhysB50.135301}%
  \BibitemOpen
  \bibfield  {author} {\bibinfo {author} {\bibfnamefont {Z.~B.}\ \bibnamefont
  {Li}}, \bibinfo {author} {\bibfnamefont {Y.~M.}\ \bibnamefont {Liu}},
  \bibinfo {author} {\bibfnamefont {D.~X.}\ \bibnamefont {Yao}}, \ and\
  \bibinfo {author} {\bibfnamefont {C.~G.}\ \bibnamefont {Bao}},\ }\href
  {\doibase 10.1088/1361-6455/aa7440} {\bibfield  {journal} {\bibinfo
  {journal} {J. Phys. B: At. Mol. Opt. Phys.}\ }\textbf {\bibinfo {volume}
  {50}},\ \bibinfo {pages} {135301} (\bibinfo {year} {2017})}\BibitemShut
  {NoStop}%
\bibitem [{\citenamefont {He}\ \emph {et~al.}(2017)\citenamefont {He},
  \citenamefont {Liu},\ and\ \citenamefont
  {Bao}}]{He2017.CommunTheorPhys68.220}%
  \BibitemOpen
  \bibfield  {author} {\bibinfo {author} {\bibfnamefont {Y.-Z.}\ \bibnamefont
  {He}}, \bibinfo {author} {\bibfnamefont {Y.-M.}\ \bibnamefont {Liu}}, \ and\
  \bibinfo {author} {\bibfnamefont {C.-G.}\ \bibnamefont {Bao}},\ }\href
  {\doibase 10.1088/0253-6102/68/2/220} {\bibfield  {journal} {\bibinfo
  {journal} {Comm. Theor. Phys.}\ }\textbf {\bibinfo {volume} {68}},\ \bibinfo
  {pages} {220} (\bibinfo {year} {2017})}\BibitemShut {NoStop}%
\bibitem [{Note2()}]{Note2}%
  \BibitemOpen
  \bibinfo {note} {Note that the system satisfies $c_{AB}^2<c_{A}c_{B}$ and is
  in the miscible regime.}\BibitemShut {Stop}%
\bibitem [{\citenamefont {Luo}\ \emph {et~al.}(2007)\citenamefont {Luo},
  \citenamefont {Li},\ and\ \citenamefont {Bao}}]{Luo2007.PRA75.043609}%
  \BibitemOpen
  \bibfield  {author} {\bibinfo {author} {\bibfnamefont {M.}~\bibnamefont
  {Luo}}, \bibinfo {author} {\bibfnamefont {Z.}~\bibnamefont {Li}}, \ and\
  \bibinfo {author} {\bibfnamefont {C.}~\bibnamefont {Bao}},\ }\href {\doibase
  10.1103/PhysRevA.75.043609} {\bibfield  {journal} {\bibinfo  {journal} {Phys.
  Rev. A}\ }\textbf {\bibinfo {volume} {75}},\ \bibinfo {pages} {043609}
  (\bibinfo {year} {2007})}\BibitemShut {NoStop}%
\bibitem [{\citenamefont {Luo}\ \emph {et~al.}(2008)\citenamefont {Luo},
  \citenamefont {Bao},\ and\ \citenamefont {Li}}]{Luo2008.JPhysB41.245301}%
  \BibitemOpen
  \bibfield  {author} {\bibinfo {author} {\bibfnamefont {M.}~\bibnamefont
  {Luo}}, \bibinfo {author} {\bibfnamefont {C.}~\bibnamefont {Bao}}, \ and\
  \bibinfo {author} {\bibfnamefont {Z.}~\bibnamefont {Li}},\ }\href {\doibase
  10.1088/0953-4075/41/24/245301} {\bibfield  {journal} {\bibinfo  {journal}
  {J. Phys. B: At. Mol. Opt. Phys.}\ }\textbf {\bibinfo {volume} {41}},\
  \bibinfo {pages} {245301} (\bibinfo {year} {2008})}\BibitemShut {NoStop}%
\bibitem [{\citenamefont {Xu}\ \emph {et~al.}(2009)\citenamefont {Xu},
  \citenamefont {Zhang},\ and\ \citenamefont {You}}]{Xu2009.PRA79.023613}%
  \BibitemOpen
  \bibfield  {author} {\bibinfo {author} {\bibfnamefont {Z.~F.}\ \bibnamefont
  {Xu}}, \bibinfo {author} {\bibfnamefont {Y.}~\bibnamefont {Zhang}}, \ and\
  \bibinfo {author} {\bibfnamefont {L.}~\bibnamefont {You}},\ }\href {\doibase
  10.1103/PhysRevA.79.023613} {\bibfield  {journal} {\bibinfo  {journal} {Phys.
  Rev. A}\ }\textbf {\bibinfo {volume} {79}},\ \bibinfo {pages} {023613}
  (\bibinfo {year} {2009})}\BibitemShut {NoStop}%
\bibitem [{\citenamefont {Shi}\ and\ \citenamefont
  {Ge}(2011)}]{Shi2011.PRA83.013616}%
  \BibitemOpen
  \bibfield  {author} {\bibinfo {author} {\bibfnamefont {Y.}~\bibnamefont
  {Shi}}\ and\ \bibinfo {author} {\bibfnamefont {L.}~\bibnamefont {Ge}},\
  }\href {\doibase 10.1103/PhysRevA.83.013616} {\bibfield  {journal} {\bibinfo
  {journal} {Phys. Rev. A}\ }\textbf {\bibinfo {volume} {83}},\ \bibinfo
  {pages} {013616} (\bibinfo {year} {2011})}\BibitemShut {NoStop}%
\end{thebibliography}%
\end{document}